\documentclass[twocolumn]{aastex63}

\received{\today}
\revised{\today}
\accepted{\today}
\shorttitle{StelNet}
\shortauthors{Garraffo et al.}

\usepackage{todonotes}
\usepackage{amsmath}	% Advanced maths commands
\usepackage{amssymb}	% Extra maths symbols
\usepackage{breqn} %break maths lines up 
\usepackage{hyperref}

\begin{document}
\title{StelNet: Hierarchical Neural Network for Automatic Inference in Stellar Characterization}

\correspondingauthor{Cecilia Garraffo}
\email{cgarraffo@cfa.harvard.edu}

\author[0000-0002-8791-6286]{Cecilia Garraffo}
\affiliation{Institute for Applied Computational Science, Harvard University, 33 Oxford St., Cambridge, MA 02138, USA}
\affiliation{Harvard-Smithsonian Center for Astrophysics, 60 Garden St., Cambridge, MA 02138, USA}

\author[0000-0002-8791-6286]{Pavlos Protopapas}
\affiliation{Institute for Applied Computational Science, Harvard University, 33 Oxford St., Cambridge, MA 02138, USA}

\author[0000-0002-0210-2276]{Jeremy J. Drake}
\affiliation{Harvard-Smithsonian Center for Astrophysics, 60 Garden St., Cambridge, MA 02138, USA}

\author[0000-0002-8358-0295]{Ignacio Becker}
\affiliation{Departamento de Ciencia de la Computación, Facultad de Ingeniería, Pontificia Universidad Católica de Chile, Av. Vicuña Mackenna 4860, 7820436 Macul, Santiago}
\affiliation{Millennium Institute of Astrophysics, Faculty of Physics, Av. Vicuña Mackenna 4860, 7820436 Macul, Santiago, Chile}

\author[0000-0002-8791-6286]{Phillip Cargile}
\affiliation{Harvard-Smithsonian Center for Astrophysics, 60 Garden St., Cambridge, MA 02138, USA}

%%%%%%%%%%%%%%%%%%%%%%%%%%%%%%%%%%%%%%%%%%%%%%%%%%%%%%%%%%%%%%%%%%%%%%%%%%%%%%%%%%%%%
% Abstract
%%%%%%%%%%%%%%%%%%%%%%%%%%%%%%%%%%%%%%%%%%%%%%%%%%%%%%%%%%%%%%%%%%%%%%%%%%%%%%%%%%%%%

\begin{abstract}

Characterizing the fundamental parameters of stars from observations is crucial for studying the stars themselves, their planets, and the galaxy as a whole. 
Stellar evolution theory predicting the properties of stars as a function of stellar age and mass enables translating observables into physical stellar parameters by fitting the observed data to synthetic isochrones.  However, the complexity of overlapping evolutionary tracks often makes this task numerically challenging, and with a precision that can be highly variable, depending on the area of the parameter space the observation lies in.  
This work presents {\sc StelNet}, a Deep Neural Network trained on stellar evolutionary tracks that quickly and accurately predicts mass and age from absolute luminosity and effective temperature for stars with close to solar metallicity. The underlying model makes no assumption on the evolutionary stage and includes the pre-main sequence phase. We use bootstrapping and train many models to quantify the uncertainty of the model. To break the model's intrinsic degeneracy resulting from overlapping evolutionary paths, we also built a hierarchical model that retrieves realistic posterior probability distributions of the stellar mass and age. We further test and train {\sc StelNet} using a sample of stars with well-determined masses and ages from the literature.

\end{abstract}

\keywords{stars: rotation --- stars: magnetic field --- stars: evolution }

\let\thefootnote\relax\footnotetext{\hspace{- 0.2 in } \textbf{Software: {\sc StelNet}}, DOI: \href{https://zenodo.org/badge/latestdoi/358709869}{358709869}, Repository: 
\href{https://github.com/cgarraffo/StelNet}{GITHUB}.}
\let\thefootnote\relax\footnotetext{\hspace{- 0.2 in } License: authors retain copyright and release the work under} 
\let\thefootnote\relax\footnotetext{\hspace{- 0.2 in } The Open Source MIT License. If {\sc StelNet} is useful for your} 
\let\thefootnote\relax\footnotetext{\hspace{- 0.2 in }  research, please cite this paper in your work. 
}

%%%%%%%%%%%%%%%%%%%%%%%%%%%%%%%%%%%%%%%%%%%%%%%%%%%%%%%%%%%%%%%%%%%%%%%%%%%%%%%%%%%%%
% Introduction
%%%%%%%%%%%%%%%%%%%%%%%%%%%%%%%%%%%%%%%%%%%%%%%%%%%%%%%%%%%%%%%%%%%%%%%%%%%%%%%%%%%%%

%@arxiver{MixtureModels.pdf, posterior_distr_age_gaia_i=55.eps, Tension_data_TL.eps}

\section{INTRODUCTION}
\label{sec:Intro}

Stellar evolution theory allows us to predict the luminosity and radius of a star of a given initial mass and metallicity as a function of age. Under the black body approximation, it also predicts stellar effective temperature. However, astronomers usually deal with the converse problem of retrieving fundamental stellar parameters from observable quantities like apparent magnitude and photometric color.  This problem is often non-trivial because stellar evolutionary tracks are not of functional form and sometimes cross each other in the Hertzprung-Russell parameter space leading to degeneracies. In addition, the difficulty of fitting an observation to evolutionary tracks and the probabilistic security of the result strongly depend on the region of the parameter space of the observation in question.

To estimate stellar masses and ages, one usually needs first to determine the distance and then use interpolation within model stellar evolutionary tracks or isochrones \citep[e.g.][]{ Jorgensen.Lindegren:05,  Juric.etal:08, Breddels.etal:10, Burnett.Binney:10, Binney.etal:14, Morton:15}.
\citet{Pont.Eyer:04} and \citet{Takeda.etal:07} pointed out
that interpolating between evolutionary tracks or isochrones does not account for the nonlinear mapping of time onto the H–R diagram, and neither for the non-uniform stellar mass distribution observed in the galaxy. This introduces a bias toward older derived ages than expected from stellar evolution theory.

Neural networks (NNs) are more flexible compared to classical interpolation methods, and they are also much faster at making predictions once trained. They become especially convenient when a large number of predictions are expected from a model trained on a single dataset.  In addition, and unlike interpolation methods where all data needs to be stored and used to make each prediction, a NN uses all training data only once. Then only the weights and biases need to be stored and used for future predictions.  Here, we train a NN on synthetic evolutionary data computed using the Modules for Experiments
in Stellar Astrophysics (MESA) suite of programs as implemented in the MESA Isochrones \& Stellar Tracks ({\it MIST}) project \citep{Dotter:16, Choi.etal:16, Paxton.etal:11, Paxton.etal:13, Paxton.etal:15} and use it to predict masses and ages of stars given their effective temperatures ($T_{eff}$) and bolometric luminosities ($L$).

The confidence levels of the evolutionary track and isochrone-based estimates of stellar parameters have historically been ignored. Such confidence levels encompass two kinds of errors of different nature: the epistemic error, that reflects the ability of the model to describe the relationship between variables, and the aleatory error, the one that arises from the limited precision of observations. 
The uncertainties of track- and isochrone-based inference are crucial for any physical interpretation. Recently, \cite{Cargile.etal:20} have quantified those for their model using mocks drawn from the model space, on a case by case basis. 

In this work, we attempt to quantify the uncertainty of a model (the epistemic error) that retrieves physical stellar parameters from close-to-observable quantities. Historically, this has been neglected, an approach justified by the fact that aleatory errors are typically much larger. However, it is important (and good practice) to have bounds for the model's reliability as a function of the parameter space.  This is especially the case when the precision of observations has increased dramatically in recent years due to advances like the {\it Gaia} mission that have reduced the aleatory error and comparatively increased the impact of the epistemic (or systematic) errors.

Quantifying uncertainty is a challenge for deep learning since any reliable method for estimating posterior probability distributions, e.g., Markov Chain Monte Carlo, scales badly with the number of parameters. As a consequence, all Bayesian Neural Networks provide approximate posteriors and, therefore, credible intervals. A reliable way of estimating the epistemic error in Deep NNs (DNNs) is to train a number of networks on bootstrapped versions of the data. It can be shown that this is equivalent to having a Bayesian Neural Network (BNN) where the priors of the model are the different initializations of the networks to be trained \citep{Osband.etal:18}.

Finally, to properly treat the degeneracies that arise from evolutionary track cross-overs, and to address the biases toward older ages introduced by simple interpolation methods, we build a hierarchical model that probabilistically combines multiple predictions, making use of the relative abundances of stellar masses and stars of different ages in the galaxy. Hierarchical probabilistic models are well studied \citep[see][and references therein, for details]{Bishop:06}  and deal with observations that belong to different populations.
 
In our case, the main degeneracy arises from the evolutionary tracks of young stars, before reaching the so-called Zero-Age Main Sequence (ZAMS), and after ZAMS. For that reason, the degeneracy in the input parameters largely breaks if we separate the data into two populations: pre- and post-ZAMS). We build a predictive model for each population and then combine them in a hierarchical way to get a prediction, both a deterministic and probabilistic one.

In this paper:
\begin{itemize}
\item We present {\sc StelNet}, a method to perform inference on the mass and age of solar metallicity stars of any age (including pre-Main Sequence) given their absolute luminosity and effective temperature, based on MESA evolutionary tracks \citep{Paxton.etal:11, Paxton.etal:15} which is: 

\begin{itemize}
\item Accurate: provides accurate predictions for both stellar mass and age.
\item Computationally Efficient: significantly reduces the run time to perform inference in large data sets by circumventing the need to perform interpolation for each observation and the use of full probabilistic methods such as MCMC. 
\item Advantageous for Low Storage and High Re-usability: significantly reduces the  amount of synthetic data storage needed for interpolation methods.
\item Able to Quantify Uncertainty: a model with quantified epistemic uncertainty as a function of the parameter space.
\end{itemize}
\item We developed a hierarchical NN for inference and a mixture of models to properly treat the degeneracies of the data.
\item We obtain realistic posterior probability distributions on the output parameters. 

\end{itemize}

\section{Background and Previous Work}

Stellar evolution theory allows us to generate synthetic evolutionary tracks (such as MESA tracks and isochrones) that predict the physical parameters of stars of a given initial mass and metallicity throughout their life.  Traditionally, these tracks are interpolated to derive fundamental properties from the observed quantities.  Interpolation methods are susceptible to grid size and discretization. Therefore, the amount of storage required will increase with the size of the grid, as will the computational time of the interpolation \citep[see, for example][]{Morton:15}. While there are work arounds for this \citep[see, for example][]{Cargile.etal:20}, these generally involve some form of simplification of the interpolation method and are more a tool of necessity than desirability.

\subsection{MIST Data}
\label{sec:MIST}

\begin{figure}[htp]

\center
\includegraphics[trim = .in .in  .0in 0.in,clip, width = 0.49\textwidth]{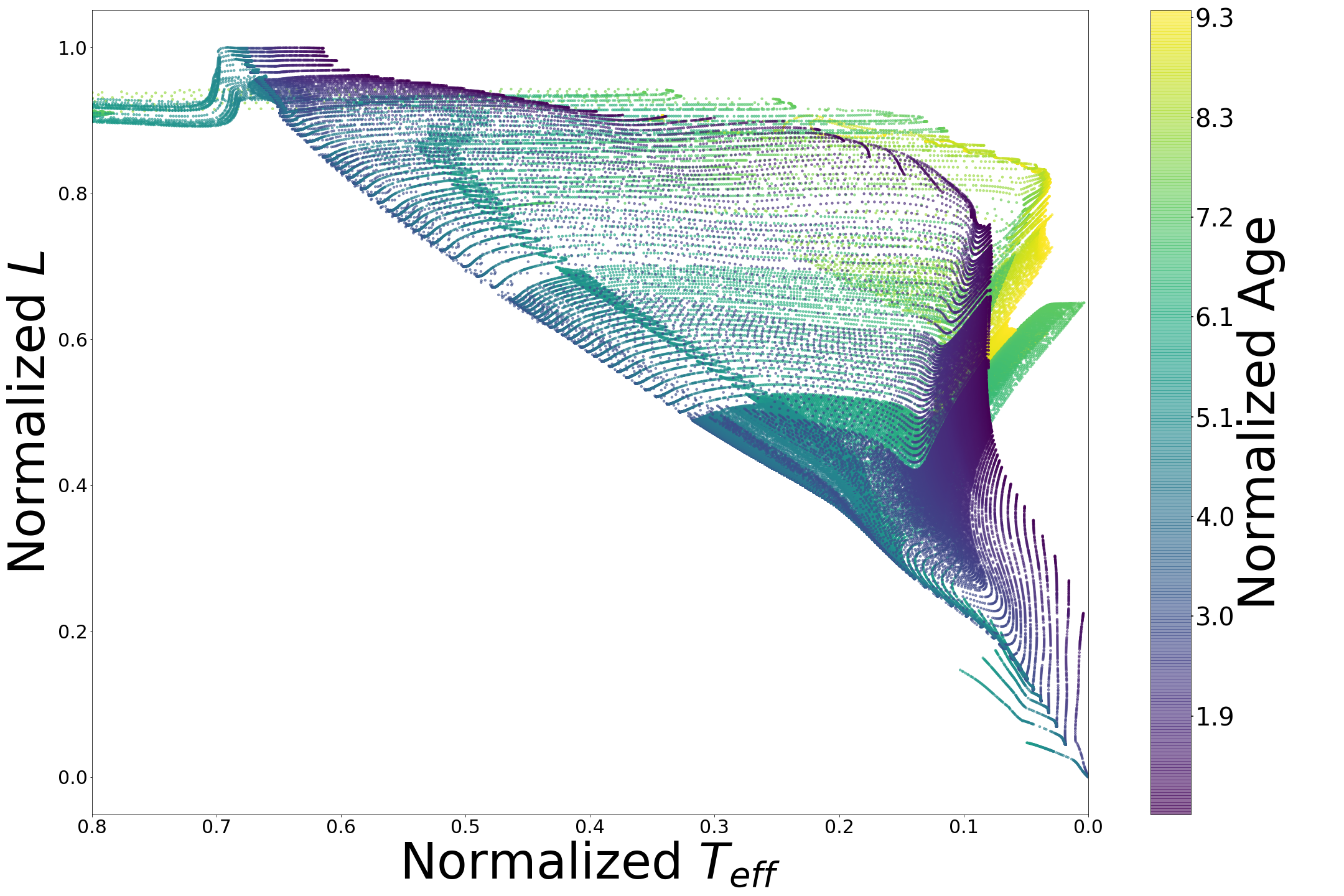}
\caption{\textit{MIST} isochrones of stars with masses between $0.01 M_{\odot}$ and $300 M_{\odot}$ with grid spacing of $0.1 M_{\odot}$ for stars with masses $M_{\odot}<3 $, $0.2 M_{\odot}$ for $3<M_{\odot}<8$, $1 M_{\odot}$ for $8<M_{\odot}<20$, $2 M_{\odot}$ for $20<M_{\odot}<40$, $5 M_{\odot}$ for $40<M_{\odot}<150$, and  $50 M_{\odot}$ for $150<M_{\odot}<300$.}% 
\label{fig:2d_data}
\end{figure}

We take advantage of this potentially unlimited amount of synthetic data and train models to learn the relationship between stellar mass and age and their $T_{eff}$ and $L$. The goal is to reverse it to retrieve the fundamental properties of stellar mass and age. We use {\it MIST} evolutionary tracks  \citep{Dotter:16,Choi.etal:16}, computed with the MESA codes, due to their extensive grid and ready public availability.\footnote{http://waps.cfa.harvard.edu/MIST/}  Figure~\ref{fig:2d_data} shows the evolutionary tracks for a grid of stellar masses between $0.01 M_{\odot}$ and $300 M_{\odot}$  with spacing that depends on the mass. The axes represent variables linked to observables ($L$ and $T_{eff}$), plus the Equivalent Evolutionary Point (EEP) that is a measure of age \citep{Paxton.etal:11}.  We plot this last one to illustrate the degeneracy of age and mass from the observed quantities and unfold it.

\subsection{Methods}

Building a mathematical model is particularly challenging when dealing with degeneracies.  A way to deal with this is through so-called hierarchical models that separate populations to break the degeneracy and then combine predictions with a mixture of distributions. 

Neural Networks (NN) have proven to be very powerful tools for accurate predictions of complex problems, providing a framework that results in impressive performance \citep[see, for example, ][and references therein]{LeCun.etal:15}. However, deep learning faces the challenge of uncertainty quantification. For most real-world applications, a prediction gains more meaning when accompanied by an uncertainty level. This is crucial when models are used to make decisions since the uncertainty will translate into risk.

Traditional neural networks use fixed weights and biases that determine how an input is transformed into an output and, therefore, are unable to quantify the confidence of their estimations. Instead, Bayesian Neural Networks (BNN) use probability distributions as their weights and biases. An estimate is obtained making multiple runs (forward passes) of the network, each time with a new set of sampled weights and biases. Each output has now multiple values instead of a single estimate. This set of output values represents a probability distribution that can be interpreted as a point estimate and an uncertainty for each of the outputs. 

Markov-chain Monte Carlo (MCMC) is one of the most robust techniques for finding posterior distributions, since it guarantees convergence. However, NNs parameter space's high dimensionality and the strong scaling of MCMC run-times with the number of free parameters makes it challenging (when not impossible) to find these posteriors on weights and biases.  For this reason, there has been a lot of development in alternative methods for BNNs.  However, these are all approximations with different levels of success. 

Hamiltonian Monte Carl \citep[HMC;][]{Neal:12}, which is de facto MCMC, methods are believed to be accurate for finding posterior distributions on BNNs parameters. Unfortunately, they scale poorly with parameter dimensionality and dataset sizes \citep{Welling.Teh:11, Chen.etal:14}. Stochastic Gradient Langevin Dynamics (SGLD) and Stochastic Gradient HMC can fix this dependency but assume a good estimation of the gradient noise. The need to address this issue resulted in a lot of interesting new approaches.  
  
The mainstream variational approximation to MCMC for models with a high number of parameters is Mean Field Variational Inference \citep[MFVI;][]{Graves:11, Blundell.etal:15}. However, it is known that it can severely underestimate uncertainties and (and because) it neglects the correlation of the posterior parameters \citep{Giordano.etal:15}. These shortcomings have resulted in many proposed alternatives.  In particular, some methods were designed to capture the correlation of parameters in the posterior probabilities \citep{Louizos.Welling:16, Pawlowski.etal:17, Hernandez-Lobato.Adams:15, Hernandez-Lobato.etal:16}.  The most salient of these methods were recently compared by \cite{Yao.etal:19}. They generate a simple synthetic data set, use HMC as the ground truth, and they find all of these methods to suffer from the same overconfidence in regions with no training data.  

The Linear Response (LR) Methods for Accurate Covariance Estimates from Mean Field Variational Bayes  \citep[MFVB;][]{Giordano.etal:15} and the LR-GLM: High-Dimensional Bayesian Inference Using LR Data Approximations \citep{Trippe.etal:19} preprocess or summarize data sets and provide posteriors very similar to the ones by MCMC. However, these methods are applicable only when the posterior approximation is in the exponential family. Dropout as a Bayesian Approximation \citep{Gal.Ghahramani:16} is another method that provides posterior distributions but is mathematically equivalent to MFVB and, therefore, we expect it to inherit the same shortcomings. 

Finally, some methods focus on assessing and calibrating MFVB. For example, Practical Posterior Error Bounds from Variational Objectives \citep{Huggins.etal:20} provides efficient error bounds on the quality of the posterior, which useful diagnostic on the quality of a given MFVB output.  \cite{Prado.etal:19}  propose to use a dual neural network architecture to calibrate their BNN, that can be useful for problems that have data for all regions of the input parameters. A similar approach has been recently published by \cite{Amini.etal:19}.

In sight of these limitations, Variational Inference becomes a useful tool since it provides an approximation to find the closest to real posterior probability distributions from a certain family of functions, using the non-symmetric Kullback-Leibler (KL) divergence.   The mean-field approximation is the simplifying assumption that variables are independent.  MFVB is known for severely underestimating variances even for even simple multivariate Gaussian distributions.  The reason is that the MFVB approximate normal will lie inside the real posterior equating the standard deviation for all variables. As a consequence, we would expect this effect to get exacerbated in Deep Neural Networks (DNNs) due to the large number of interdependent parameters.

\begin{figure}[htp]
\center
\includegraphics[trim = .2in 0.3in  0.1in 0.in,clip, width = 0.45\textwidth]{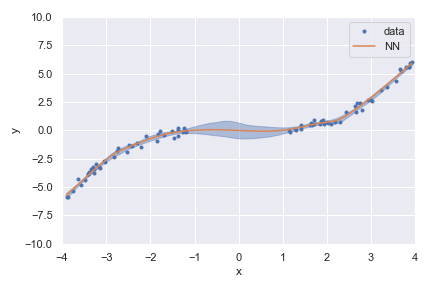}
\includegraphics[trim = .0in 0.in  0.38in 0.2in,clip, width = 0.47\textwidth]{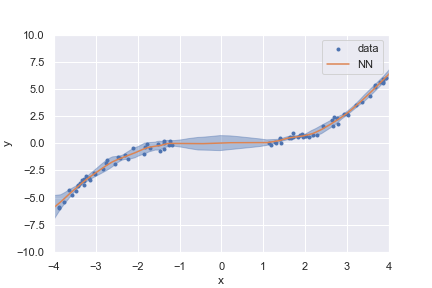}
\caption{Data generated by $y = 0.1 x^3 + \epsilon$, where $\epsilon \sim \mathcal{N}(0, 0.25)$.  Evaluated on 80 training inputs, 20 evaluation inputs uniformly sampled from $[ -4, -1 ] \cup [1, 4]$ and 200 test inputs uniformly sampled from $[ -4, 4]$. Top: MCMC - bottom: Bootstrapping, 20 models, main and 90\% of model predictions for a model with just one layer}
\label{fig:bs}
\end{figure}

Bootstrapping is another approach to quantifying a model's  epistemic uncertainty; in particular, it is very convenient for DNNs since it circumvents the issue of large runtimes of methods like MCMC. Predictive uncertainty is estimated by training multiple models on the same dataset.  The prediction of the final model is the mean of the predictions of the multiple models, and the standard deviation is the area that contains 95\% of the models.  Using a simple model, we show that the results of this bootstrapping technique are consistent with the ones of MCMC (Figure~\ref{fig:bs}).

\section{{\sc StelNet}}
\label{sec:na}
We build a probabilistic hierarchical model using bootstrapping of a mixture of models. To do that, we start with a deterministic model (Section~\ref{sec:Det}), composed of two different NNs. We then combine the results using a mixture of models (Section~\ref{sec:Mix}) that takes into account the stellar abundances in the galaxy, and finally we convert them to probabilistic ones using bootstrapping (Section~\ref{sec:BS}).  For each of the two NNs, we train a number of models using  different bootstrap'ed samples (with replacement) of the same data set and different, random, initializations.  We then do forward runs with our data through all trained models. For each NN, that provides a predictive distribution from which an expectation value and an uncertainty level can be estimated, and even the shape of this distribution, provided we have enough models trained.  

\subsection{Deterministic Model}
\label{sec:Det}

From the {\it MIST} data described in Section~\ref{sec:MIST}, it becomes clear that there is a degeneracy that cannot be circumvented in the two-dimensional $\log(L/L_{\odot})$ - $log(T_{eff}[K])$  parameter space (see Figure~\ref{fig:2d_data}).  Stars evolve with time towards lower luminosities and hotter effective temperatures until they reach ZAMS. Their luminosity then starts to increase, and the temperature slowly decreases. Pre- and post-ZAMS tracks cross over as a consequence. To deal with this, we build two models, one for before and one for after ZAMS.  These two networks have the same architecture and activation function described in Section~\ref{sec:BS} but are trained on different sets of data.  Each one performs very well on the respective test sets. 

Each model is a fully connected neural network of 10 hidden layers with 50 nodes each, that accurately predicts stellar mass and age from $T_{eff}$ and $L$ for stars of solar metallicity. Since NN are shown to be universal approximations, one increases the complexity of the model by increasing the number of layers and the number of nodes until overfitting becomes a problem. We empirically found this architecture to produce good results. 

We chose ReLU as the activation function because it yields easily calculated derivatives which we use for training and avoid vanishing gradients, and it is well established as a good activation function.  We use mean square error (MSE) as the loss function and Adam as the optimizer with a learning rate of $10^{-3}$. The rest of the parameters of the optimizer are the default values. The coefficients $\beta$, used to compute running averages of gradient and its square, are (0.9,0.999); $\epsilon$, the term to improve numerical stability, is $1e-8$; and we do not use regularization.

In {\it MIST}, the ZAMS is given by setting the parameter equivalent evolutionary points (EEP) to 202. So the pre-ZAMS model is trained on the fraction of the dataset with EEP $\le$ 202 and the post-ZAMS model is trained on the data with EEP $<$ 202. The input data, $\log(Teff \, [K])$ and $\log(L \, [L_{\odot}])$, is normalized to the range [0,1], using the  minimum and maximum values of these quantities in the training dataset for each model.  Those values are $\log(Teff \, [K])_{Pre} =(3.46, 4.77)$ and $\log(L \, [L_{\odot}])_{Pre} =(-3.00, 6.81)$ for the pre-ZAMS model, and $\log(Teff \, [K])_{Post} =(3.46, 5.39)$ and $\log(L \, [L_{L_{\odot}}])_{Post} =(-3.00, 6.82)$ for the post-ZAMS model. 

We train our model on {\it MIST} data (see Figure~\ref{fig:2d_data}). We randomly select 80\% of the data as the training set and use the rest for testing. MSE, when interpreted as a likelihood, assumes the error (in our case the epistemic or systematic error) is normally distributed and independent. Stars of different masses and ages evolve at different rates. Stellar lifetimes vary from a few million to tens of billions of years from the most massive down to the least massive stars. This means the errors associated with age predictions will not exactly be normal. We train our network on $\log(age)$ because this guarantees a closer to normal distribution and because it reduces the issue of different timescales involved in stellar evolution.

\subsection{Mixture of Models}
\label{sec:Mix}

\begin{figure*}
\center
\includegraphics[trim = .1in 0.in  0.1in 0in,clip, width =\linewidth]{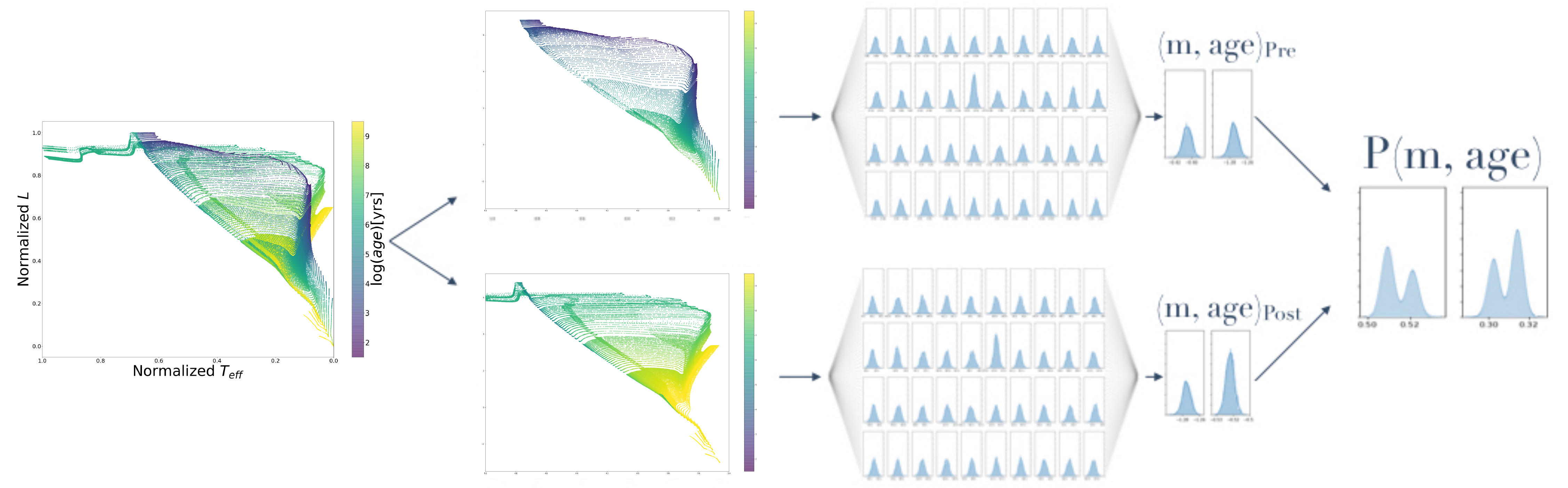}
\caption{Mixture of Inference Neural Networks diagram. From left to right: {\it MIST} synthetic data in $\log(T_{eff})$-$\log(L/L_{\odot})$ space, colored by age;  subsets of the same data, split at $EEP = 202$ which corresponds to the ZAMS; Inference Neural Networks trained on each of the subsets of the data; Gaussian posterior probabilities on mass and age for each NN, probabilistic combination of these two posteriors, using galactic abundances and IMF.}
\label{fig:mixmodel}
\end{figure*}

 While it is easy to split the synthetic data into pre- and post-ZAMS, when dealing with observations, in the absence of other diagnostics or prior information, such as proximity to young star forming regions that might strongly suggest a pre-main-sequence evolutionary phase for example, there is no way of knowing a priori which stage a star might be in.  We use a mixture of models to deal with this problem (see Figure~\ref{fig:mixmodel}). Given an observation, we use both networks to make predictions, so we get two sets of masses and ages for each observation. We validate this model on the test data by computing the errors in the prediction that lies closer to the true value. The rationale behind this is to see how good our best prediction is. We obtain MSE of $0.01$ for age and $0.005$ for mass predictions on normalized data. 
 
 Notice that there is no aleatory error included here and this network is intended to learn the MESA tables and circumvent the evolutionary track fitting. 
 
To combine these predictions, we assign a probability to each prediction based on the abundance of that stellar mass and age in the galaxy (IMF and time evolution) and on the possibility of observing it (luminosity function).  We use the ratio of the time a star of the predicted mass, $m$, spends before and after ZAMS to decide which model is more likely to be appropriate.  We then weight it by the initial mass function (IMF) from \cite{Kroupa:01}, $m^{-\alpha}$, with $\alpha = 0.3$ for $ m < 0.08$, $\alpha = 1.3$ for $0.08<m<0.5$, and $\alpha = 2.3$ for $m>0.5$, to account for how much more likely some stellar masses occur than others. We assign probabilities $p_{pre}$ and $p_{post}$ to our predictions from the pre- and post-ZAMS models:

\begin{align}
p_{pre} &= \frac{t_{ZAMS}(m_{pre})}{t_{lt}(m_{pre})}* m^{-\alpha(m_{pre})}\\
p_{post} &= \frac{[t_{lt}(m_{post})-t_{ZAMS}(m_{post})]}{t_{lt(m_{post})}} * m^{-\alpha(m_{post})}
\end{align}
where $t_{ZAMS}$(m) is the age of a star of mass $m$ at ZAMS, $t_{lt}$ is the lifetime of a star of mass $m$,  and $mass_{pre}$  and $mass_{post}$  are the mass predictions from the pre- and post-ZAMS models, respectively.  Thus now our predictions $y_{pre} = (age_{pre}, mass_{pre}$) and $y_{post}= (age_{post}, mass_{post}$), will have assigned probabilities given by $p_{pre}$ and $p_{post}$.

\subsection{Neural Network for Inference}
\label{sec:BNN}

Ideally, we would like to know the posterior probability distribution of the predictions instead of two single predictions (one coming from each model).  That would require lifting our models to probabilistic ones.

\subsection{Bootstrapping for Epistemic Uncertainty Quantification}
\label{sec:BS}
We train a number of models using bootstraps (sampling with replacement) of the {\it MIST} data set.  We then run our test data (forward) through all trained models. That provides a predictive distribution from which an expectation value and an uncertainty level can be estimated. Bootstrapping is usually used to avoid overfitting and lower the variance of a model. Adding a slight perturbation is another (empirical) approach to avoid overfitting and, therefore, help regularize and generalize a model. It is worth noting that our data have no aleatory error and, therefore, there is no risk of fitting the noise. The reason for using bootstrapping here is to estimate the uncertainty of the model. By taking different samples of the data and seeing where the models vary more, we can have an idea of the areas in the input parameter space where the model is robust and confident versus where it has high variance is highly dependent on the particular sample the model was trained on.

We re-sample the {\it MIST} data twenty times and train both pre- and post-MS models on each of them, using different, random, initializations.  We then run the test data forward through each of the twenty models and get as an output twenty values from which statistics we can get an approximation of the posterior probability distribution for the pre- and post- MS predictions on age and mass. 

In principle, the epistemic uncertainty could be arbitrarily reduced by increasing the density of the model grid and the rigor of the training procedure. However, in practice, our goal is more modest: to reduce the epistemic error to be sufficiently small that it does not induce any significant additional uncertainty in the derived masses and ages over and above that resulting from aleatory uncertainty---the observational uncertainties in $T_{eff}$ and $L$. We will show in Section~\ref{sec:ae} that for typical uncertainties in temperature and luminosity, our model grid and training satisfy this requirement.

\section{Evaluation}
\label{sec:valid}

  Our predictions are distributions and we want to compare them to the ground truth values. For the purpose of comparing distributions, we assume the posteriors of our two models, the pre and post-ZAMS ones, are Gaussians (before combining them into a bi-peaked distribution using a mixture of models). Our ground truth values, both synthetic and observed, have an associated error: the reported and the typical observation error propagated. If we assume these errors are normally distributed, then we can define Gaussian distributions for the ground truth.   
  To validate our model, we follow the prescription in \cite{Verde.etal:13} to compare distributions.   We calculate the {\it tension} that it is a way of finding consistency between two distributions in a Bayesian approach.
  The larger the tension, the more different the two distributions are.  Given two distributions, $P_{pred}$ - the posterior predicted by {\sc StelNet}, and $P_{true}$the Gaussian with mean being the observed value or synthetic-test-datapoint and variance being the reported error or the propagated typical observational error, we calculate the Bayesian Evidence, $E$,:

\begin{equation}
    E = \int P_{p} P_{t} \, d \,\Omega \nonumber
\end{equation}
 \noindent where $P_{p}, P_{t}$ are the distribution of the predicted and true values, respectively. If we assume that $P_{p}, P_{t}$ follow a normal distribution, $
  P_{p} = N(\mu_p, \sigma_p)$, $P_{t} = N(\mu_t, \sigma_t)$, then: 
\begin{eqnarray} 
E &=& \int  P_{p} P_{t} d \,\Omega =   \phi(\mu_p, \sigma_p, \mu_t, \sigma_t) \int N(\mu, \sigma) \, d \,\Omega  \nonumber \\ 
     &=& \phi(\mu_p, \sigma_p, \mu_t, \sigma_t). \nonumber 
 \end{eqnarray} 
 
 We use the fact that the product of two normal distributions is a normal distribution and the integral of a normal over the whole space is equal to one. The quantity of interest, $\phi$ is:  

\begin{equation}
\phi=\frac{1}{\sqrt{2 \pi}} \frac{1}{\sqrt{\sigma_{p}^2+\sigma_{t}^2}}\exp\left(\frac{1}{2}\frac{(\mu_{p}-\mu_{t})^2}{\sigma_{p}^2+\sigma_{t}^2}\right) 
\end{equation}

We use the normalized tension ($\tau$), introduced by \cite{Verde.etal:13}, as a measure of distributions dissimilarity: 
 
\begin{equation*}
    \tau = \frac{\bar{E} \rvert_{maxA=maxB}}{E}
\end{equation*}

\noindent  and use the Jeffrey's scale \citep{Jeffreys:73} to interpret the result: $\log{\tau}<1$ means the difference between the distributions is not significant, $1<\log{\tau}<2.5$  means the difference 
is substantial, strong if $2.5<\log{\tau}<5$, and highly significant if $2.5<\log{\tau}<5$.  We will consider our predictions consistent with the ground truth only in the first case, in which $\log{\tau}<1$.

\begin{figure}[htp]
\includegraphics[width=0.45\textwidth]{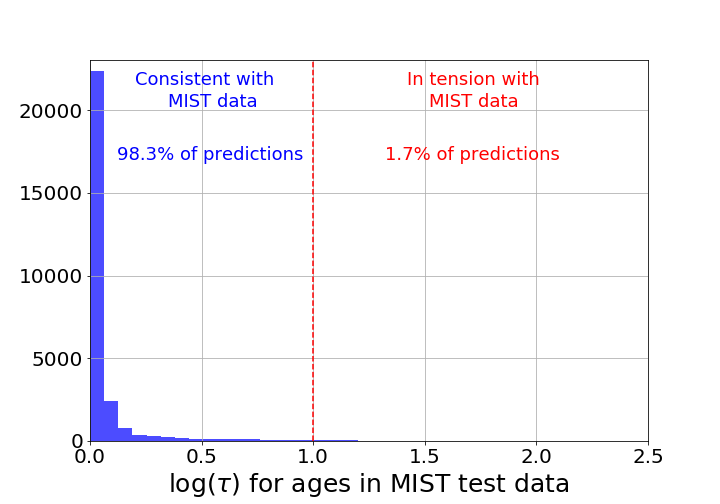}
\includegraphics[width=0.45\textwidth]{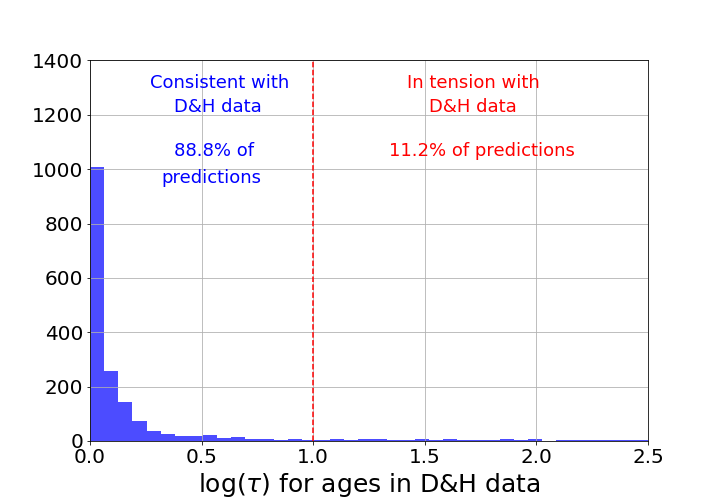}
\includegraphics[width=0.45\textwidth]{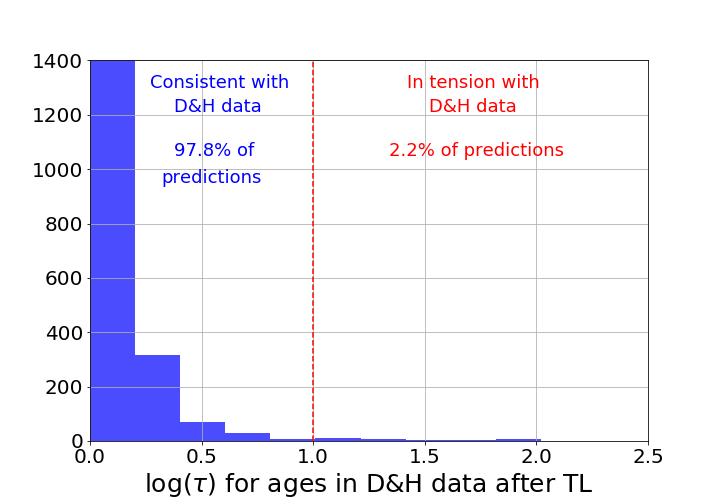}
\caption{Distribution of the log of the tension between predicted vs true values on ages for the {\it MIST} synthetic test set (upper panel), the D\&H dataset (middle panel), and the D\&H data after the networked had been trained using transfer learning (bottom panel).}
\label{fig:tension}
\end{figure}

We use a test data randomly selected from {\it MIST} synthetic isochrones, created in the same way as our training set. This data has no noise, so if our model is trained enough, it should perform well in terms of prediction.  We note that these test data (or synthetic observations) do not represent the distribution of stars in the galaxy, but instead simply reflect the time resolution in the data and the mass spacing between tracks, since we have made no attempt to selectively sample them. 
We then validate on a data set of real observations. We use a sample of field stars with estimated $T_{eff}$, $L$, $age$, and $mass$ from D\&T. Finally, we test our predictions on a well-known triple system to check if we get consistent ages for the three stars with the same age (and well-determined one) even when having different $T_{eff}$ and $L$.

\subsection{Evaluation on Synthetic Data}
\label{sec:val_mist}

\begin{figure}[htp]

\includegraphics[trim = .2in .0in  0.1in 0.1in,clip, width =0.48\textwidth]{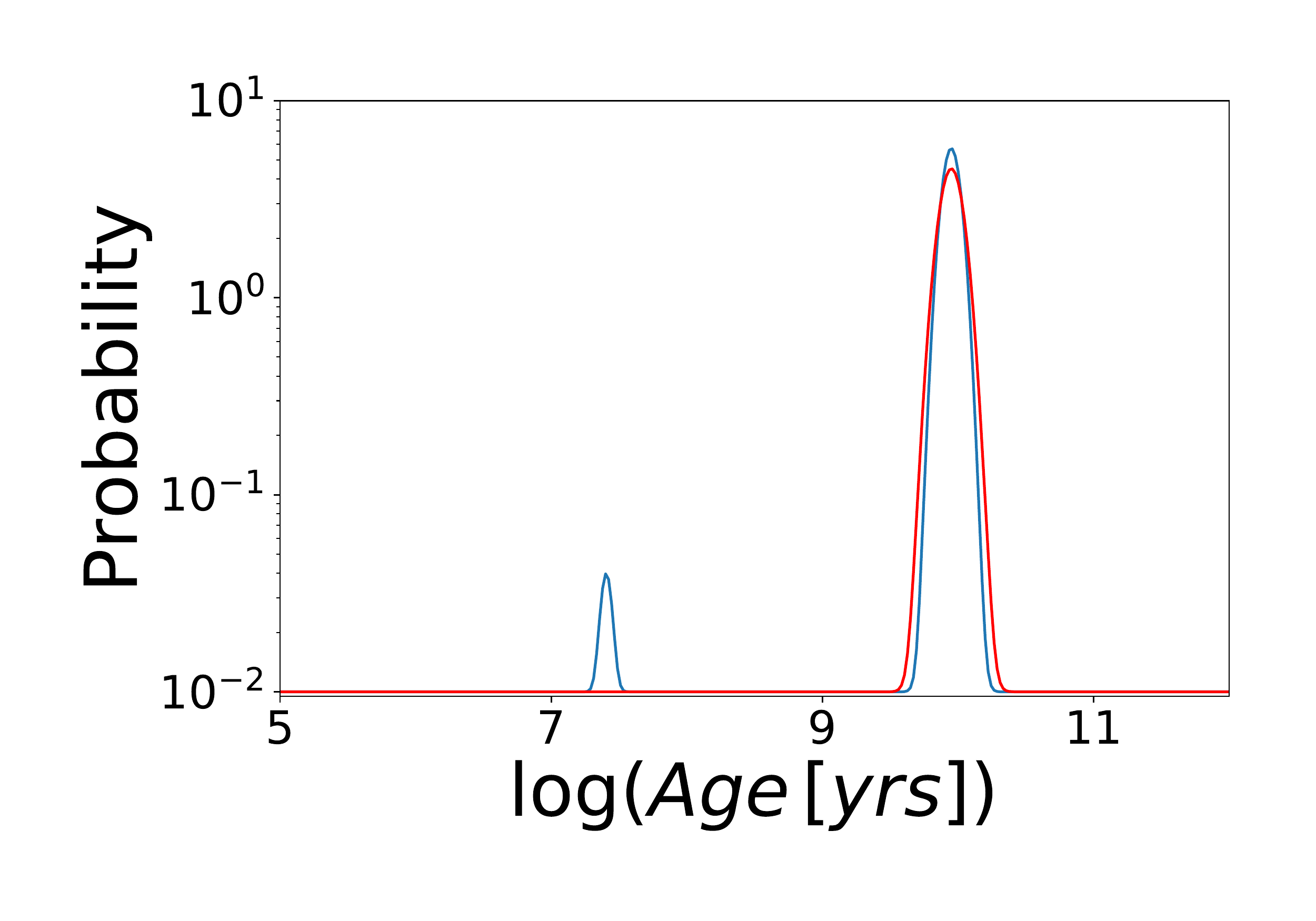}
\caption{Posterior probability distribution (blue line) for $\log(age)$ from our model for a random $0.94 M_{\odot}$ data point in the test data, and the Gaussian distribution centered on the real age from {\it MIST} and with standard deviation given by the expected propagated observational error, assuming a $50$ K temperature error in $T_{eff}$ and a a $10\%$ error in Luminosity (red line).}
\label{fig:one_post}
\end{figure}

As mentioned above, we randomly split the {\it MIST} synthetic data into a set for training and a set for testing data. We validate the model on the test set. Figure~\ref{fig:one_post} shows one prediction and its true value, while Figure~\ref{fig:post} shows the same for a number of randomly selected points in the test dataset. Notice that these distributions usually have two local maxima. The reason can be seen from the overlapping data in Figure~\ref{fig:2d_data}. For each point in the $\log(T_{eff})$ -$\log(L)$ space, there are usually two possible answers. The way we combine those two predictions was explained in detail in Section~\ref{sec:Mix} and it is based on the likelihood of observing a star of a certain mass and age. 

The {\it MIST} dataset provides one data point at each step of the evolutionary track of a star. This time step depends on how quick the evolution is at every stage and does not reflect how long a star spends in that stage.  For that reason, the relative probability of each prediction corresponding to pre- and post-MS does not match the frequency of occurrence of the test dataset.  In addition, the grid of models we use is uniformly distributed in stellar mass, which also does not reflect the initial mass function, which describes the mass dependence of star formation, and that we took into account in order to combine the two models.

We find that, for synthetic test data, 98.3 \% of age predictions and 98.7\% of mass predictions fall in the $\log{\tau}>1$ regime and, therefore, are consistent with the data (see top panel of Figure~\ref{fig:tension}) . 

\subsection{Comparison of Epistemic and Aleatory errors}
\label{sec:ae}

\begin{figure*}
\center
\includegraphics[trim = 2in 2in 2in 1in, clip, width =.7\linewidth]{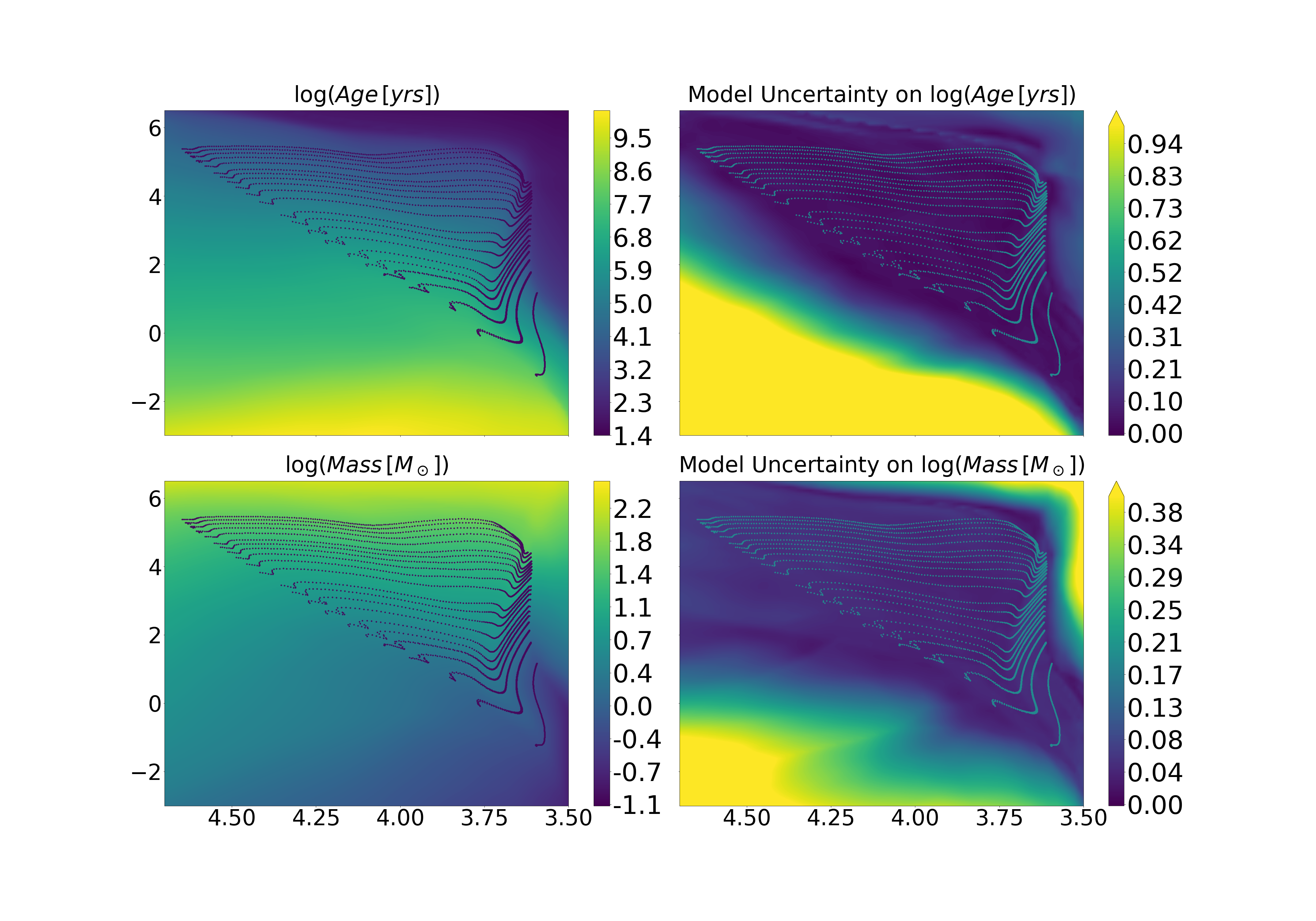}
\includegraphics[trim = 2in 2in 2in 2in, clip, width=.7\linewidth]{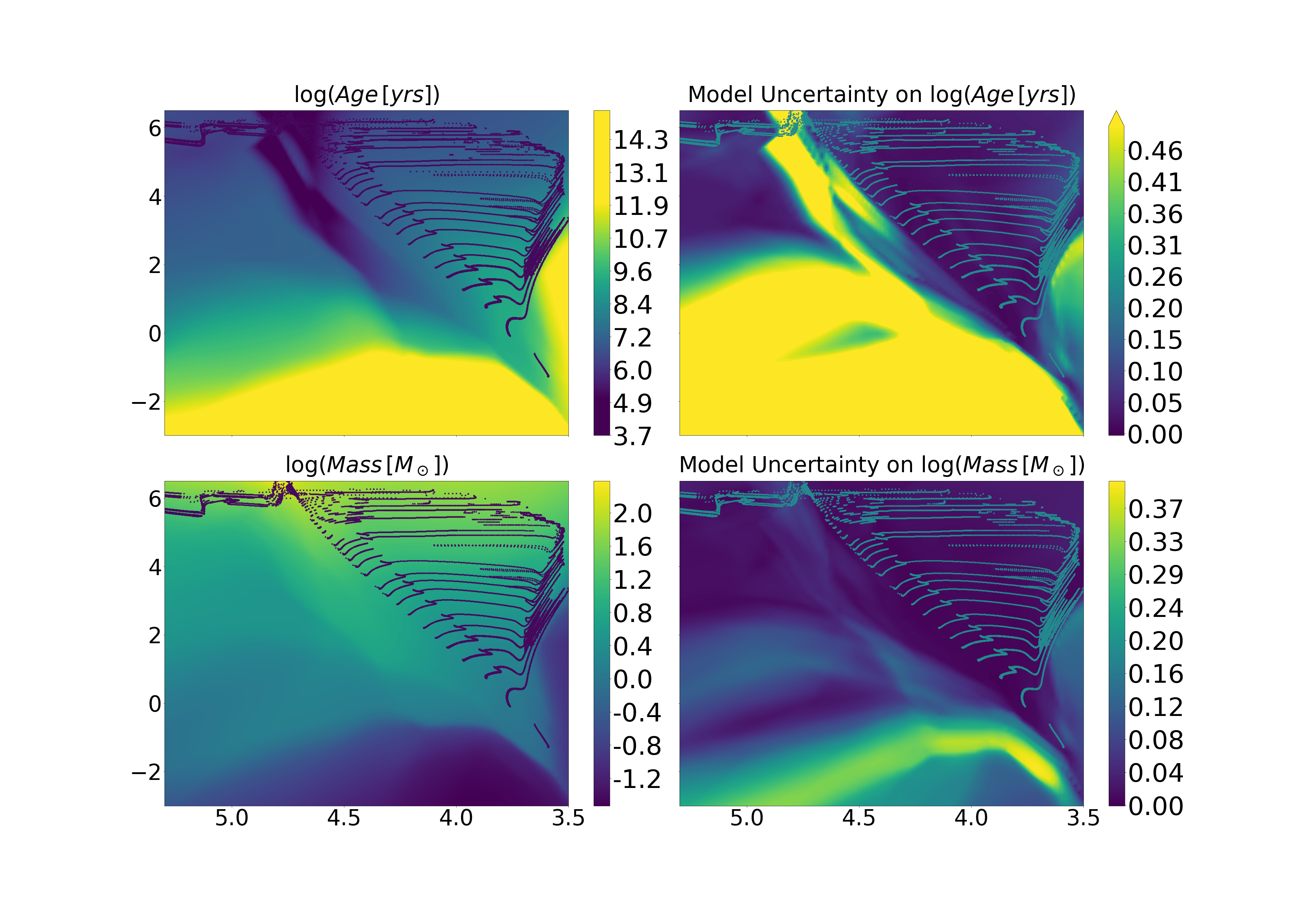}
\caption{Top four panels: mean (left) and standard deviation (right) on $log(age)$ (top) and mass (bottom) predictions for pre-ZAMS model Bottom panel: same for post-ZAMS model after using transfer learning.}
\label{fig:UQ}
\end{figure*}

\begin{figure*}
\center
\includegraphics[trim = .1in 0.1in  0.1in 0in,clip, width =.7\linewidth]{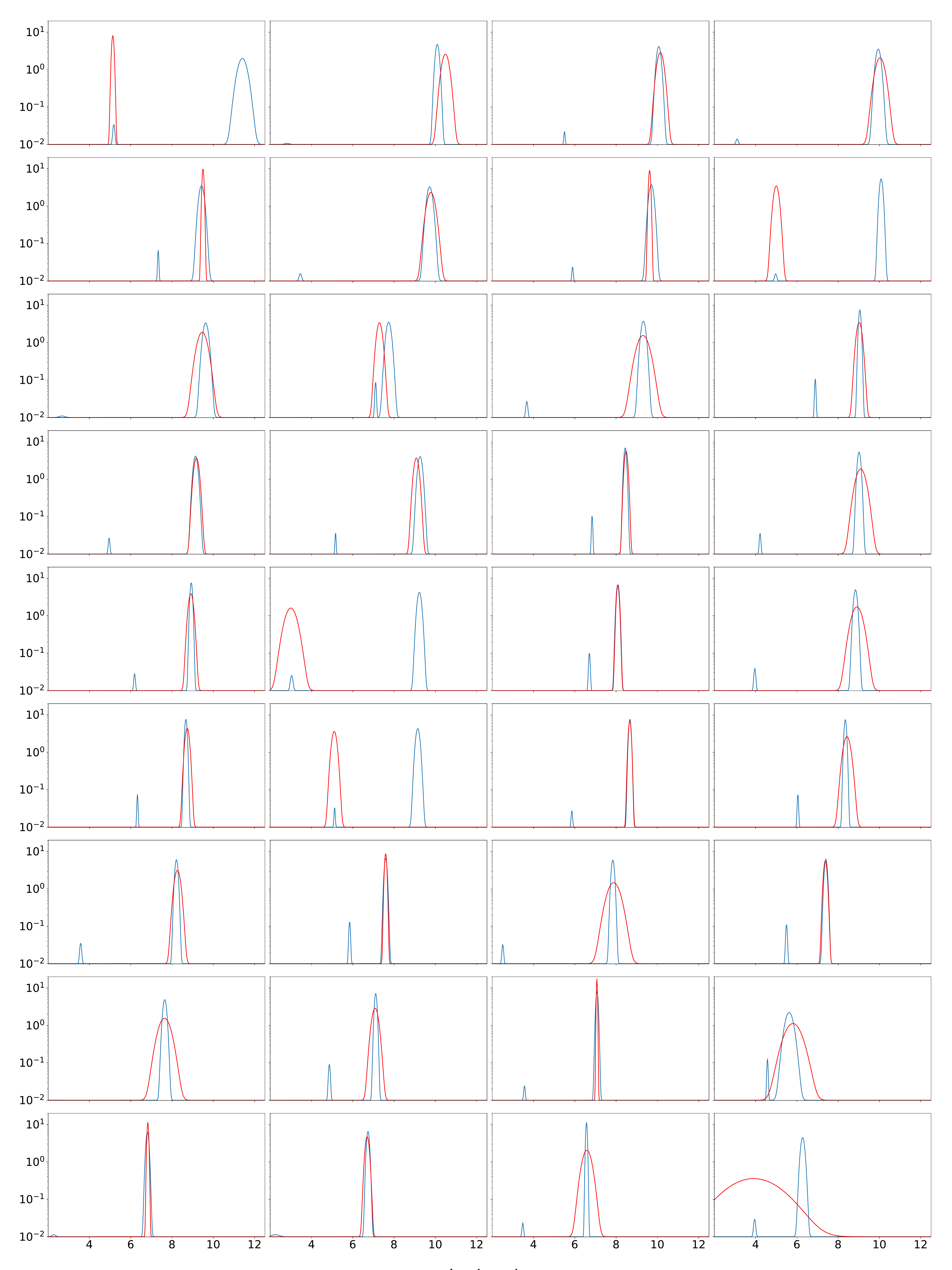}
\caption{Random selection of {\sc StelNet} posterior probability distributions on age (blue) and Gaussian representing the ground truth from {\it MIST} tables propagated with the expected observational error (red).}
\label{fig:post}
\end{figure*}

An essential advantage of having control over the epistemic error is that we can make it as small as necessary. A good metric is to compare it to the impact of the observational error on the overall uncertainty of the predictions. If our model's error is small compared to the the error introduced by the observational constraints, then we can ensure that the model error is not significantly contributing to the overall uncertainty. 
Fig.\ref{fig:post} and shows a randomly selected subset of age predictions (blue) from using {\sc StelNet} {\it MIST} test set and the true age from {\it MIST} (red). As described in \ref{sec:valid}, we assume that observations will have Gaussian noise, and we propagate a typical error on the observed quantities, $\Delta L = 0.1 L $ and $\Delta T_{eff} = 50$, to the predictions. We can then represent this ground truth by a Gaussian distribution with mean equal to the age and standard deviation equal to the propagated error. 
Figure~\ref{fig:UQ} shows our predictions of age and mass for both pre and post-ZAMS models (left panels of top and bottom plots respectively) for the whole space of parameters, as well as the standard deviations of the model (right panel). Differentiating these grids we are able to propagate the typical errors in $T_{eff}$ and $L$. We can see from Fig.\ref{fig:post} that the epistemic error (the width of the blue distribution) is typically smaller than the expected observational error. 

Moreover, when we predict on observed data (Fig~\ref{fig:post_T&H}, we find that the difference between the model's error (epistemic, blue lines) and the reported observational errors (red lines) is much larger, making our model very reliable. 

\begin{figure}[htp]
\includegraphics[width=0.48\textwidth]{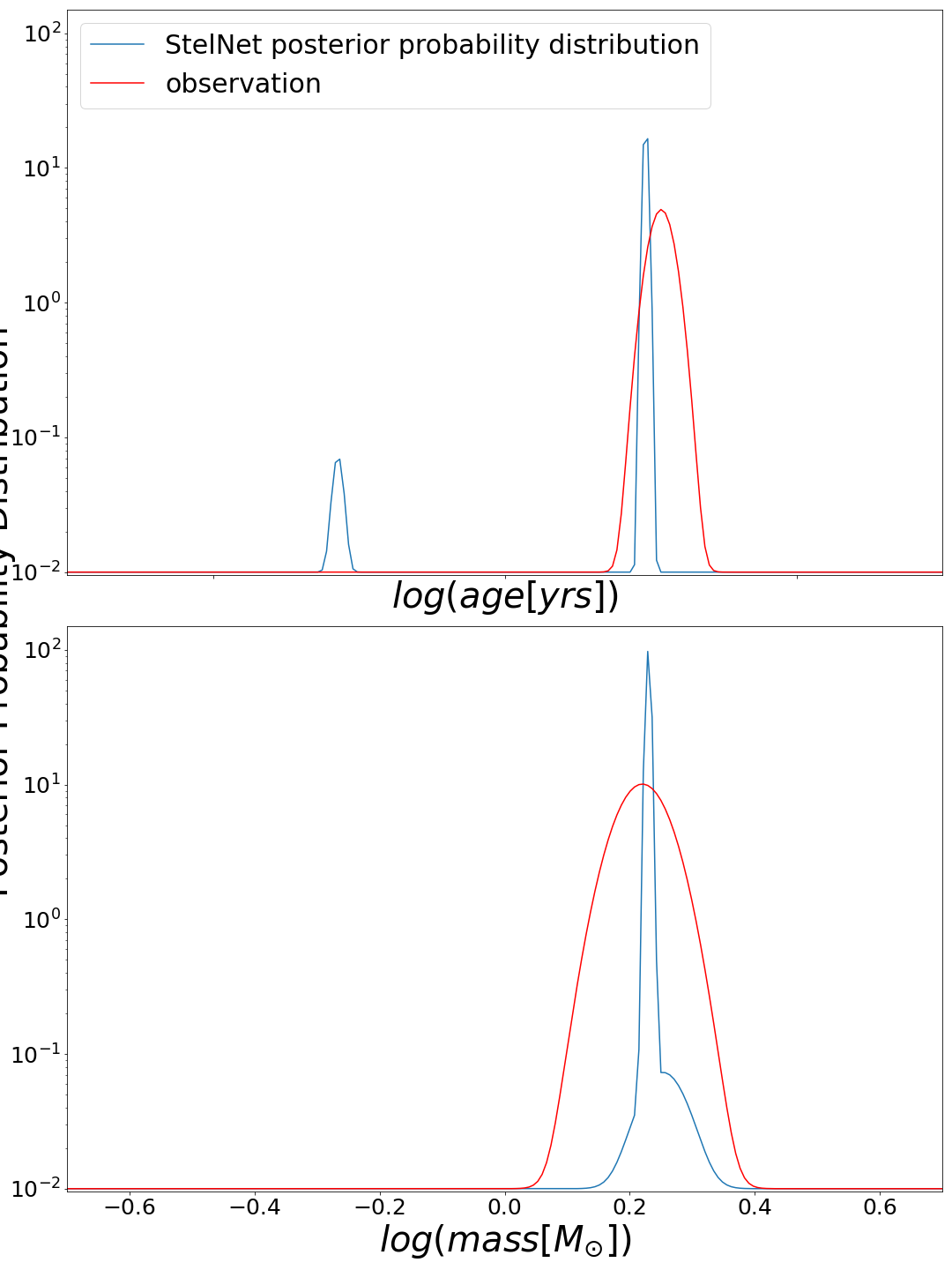}
\caption{Posterior probability distributions for age (left) and mass (right) from {\sc StelNet} (blue), and a Gaussian that represents the actual observation and error for one star (red) in the sample.}
\label{fig:one_post_gaia}
\end{figure}

\subsection{Evaluation on Observed Data}
\label{sec:val_gaia}
A more challenging test is to validate our model using real observations. So far, we have trained our models only on synthetic data, and therefore, we expect to see a difference with real observations.

\begin{figure}[htp]

\center
\includegraphics[trim = .0in 0.in  0.in .0in,clip, width=0.5\textwidth]{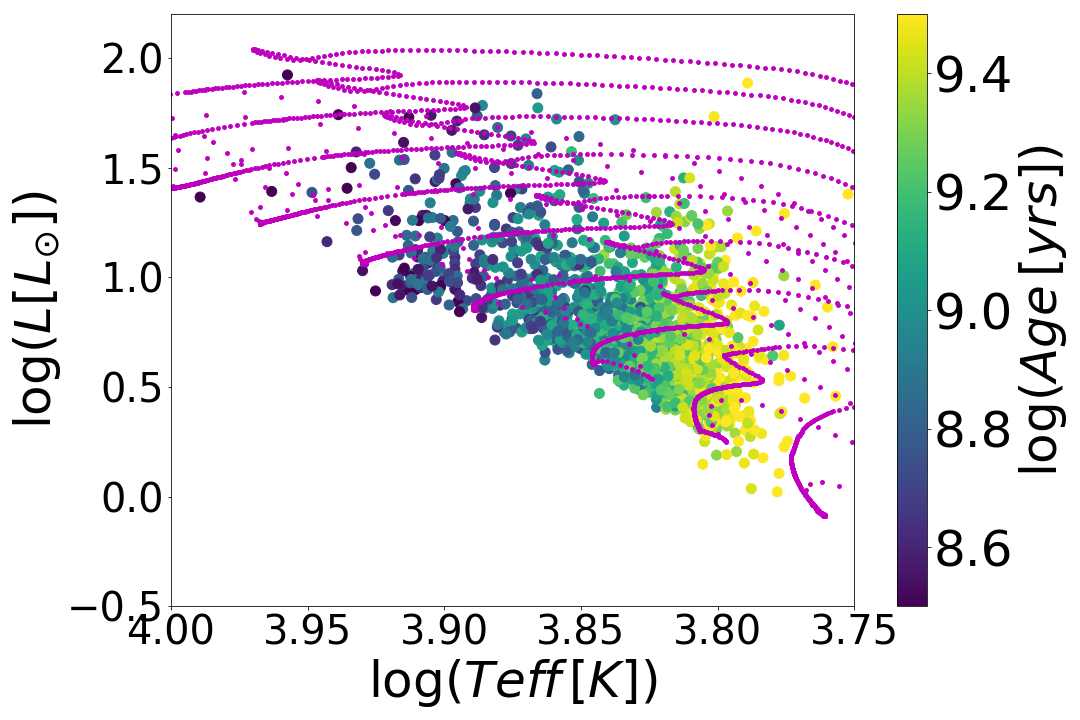}
\caption{Stellar temperatures and luminosities from the D\&H sample, colored by $log(Age)$, dot size reflects mass. A subset of {\it MIST} evolutionary tracks are overplotted in purple.}
\label{fig:TH}
\end{figure}

As a first test to see how our models perform without having seen any real observations before, we use the catalog by \citet[][hereafter D\&H]{2015ApJ...804..146D}  shown in Figure~\ref{fig:TH}.  This catalog contains measurements of the physical parameters $T_{eff}$, $mass$ and $age$ for early type (BAF) stars. In this work, authors perform a full Bayesian approach to estimate stellar masses and ages from photometry with isochrone fitting, using PARSEC solar-metallicity isochrones \citep{Bressan.etal:12}. A reliable yet much more computationally expensive method. Therefore we use it as our ground truth. We expect some systematic biases to be introduced by the the difference in isochrone data choice. 

Of the provided $age$ estimates, we used the average. This catalog uses the Hipparcos identifier which is used to match with other catalogs.  We obtain luminosity measurements from the Test Input Catalog (TIC) \citep{2019AJ....158..138S} and the Gaia DR2 stellar parameters \citep{2018A&A...616A...8A}. We prefer the TIC reported value over the Gaia one, as the former is based on spectra.

We validate the model on the observed data. Figure~\ref{fig:one_post_gaia} shows one prediction (blue) and its true value (red) for age (left) and mass (right), while Figure~\ref{fig:post_T&H} shows the same for a number of randomly selected points in the dataset.
In the {\it MIST} synthetic data, the test and training sets were generated using the same evolutionary code.  In this case, even if stellar evolution is a good predictive theory, it is never perfect, and the models will differ from the observations. We therefore, overall expect a lower accuracy in this test. 

\begin{figure}[htp]
\includegraphics[trim = .1in 0.in  0.1in .0in,clip, width =0.47\textwidth]{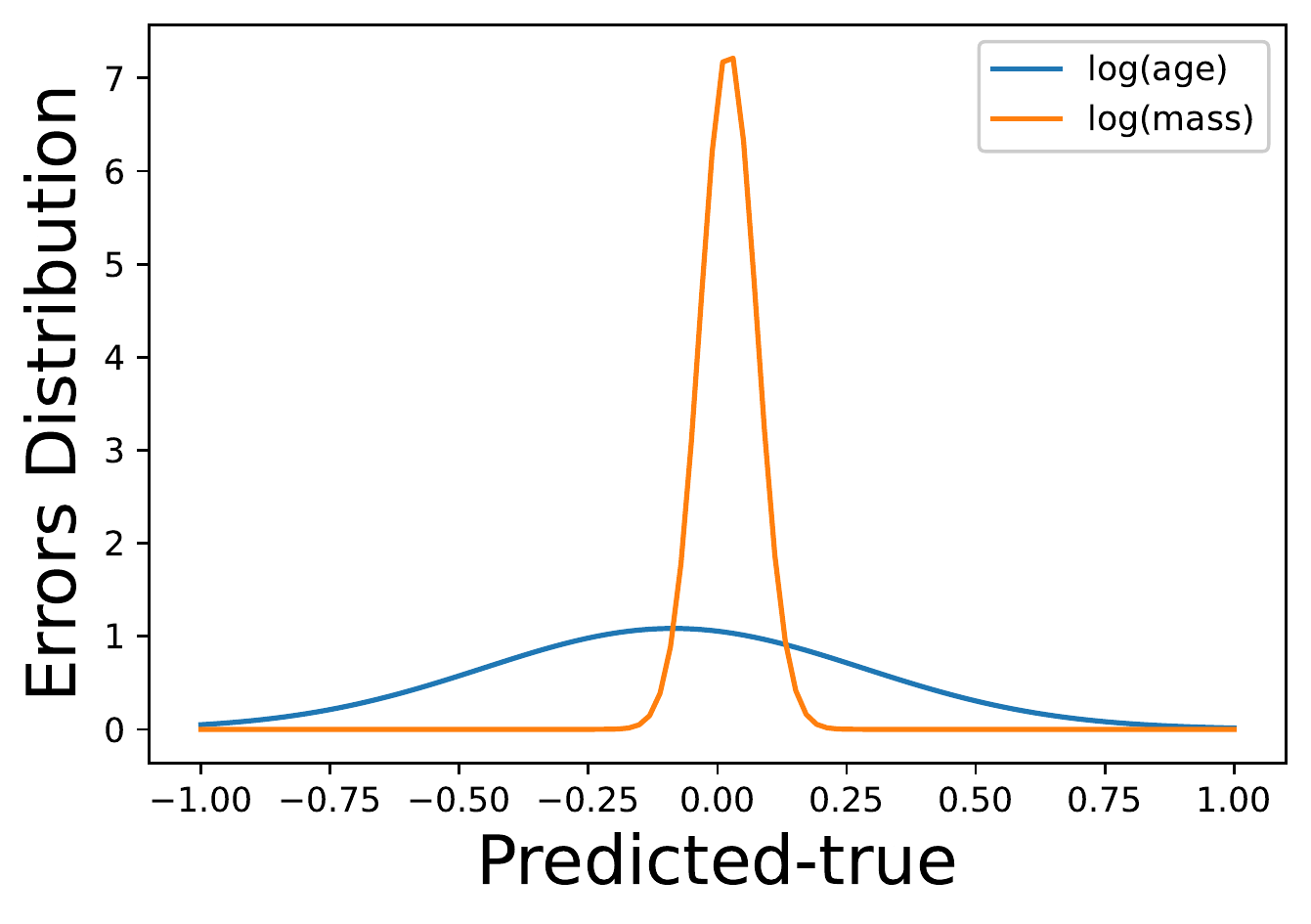}
\caption{The distributions of the $D\&H$ data prediction errors for age and mass.}
\label{fig:stats_gaia}
\end{figure}

To quantify the performance on this data set, we calculate the distribution of errors shown in Fig~\ref{fig:stats_gaia} for $\log(age)$ and $\log(mass)$. We find that, as expected, mass predictions are better than age predictions. The average error in log(age) prediction is $\sim -0.09$ with a standard deviation of $\sim 0.37$ , and the average error in $\log(mass)$ predictions are $\sim 0.02$ with a standard deviation of $\sim 0.05$. As discussed above, systematic biases are expected to result from the different set of evolutionary tracks D\&H use for their fitting. 

\begin{figure}[htp]
\includegraphics[trim = 0.1in .1in  .5in .3in,clip, width =0.49\textwidth]{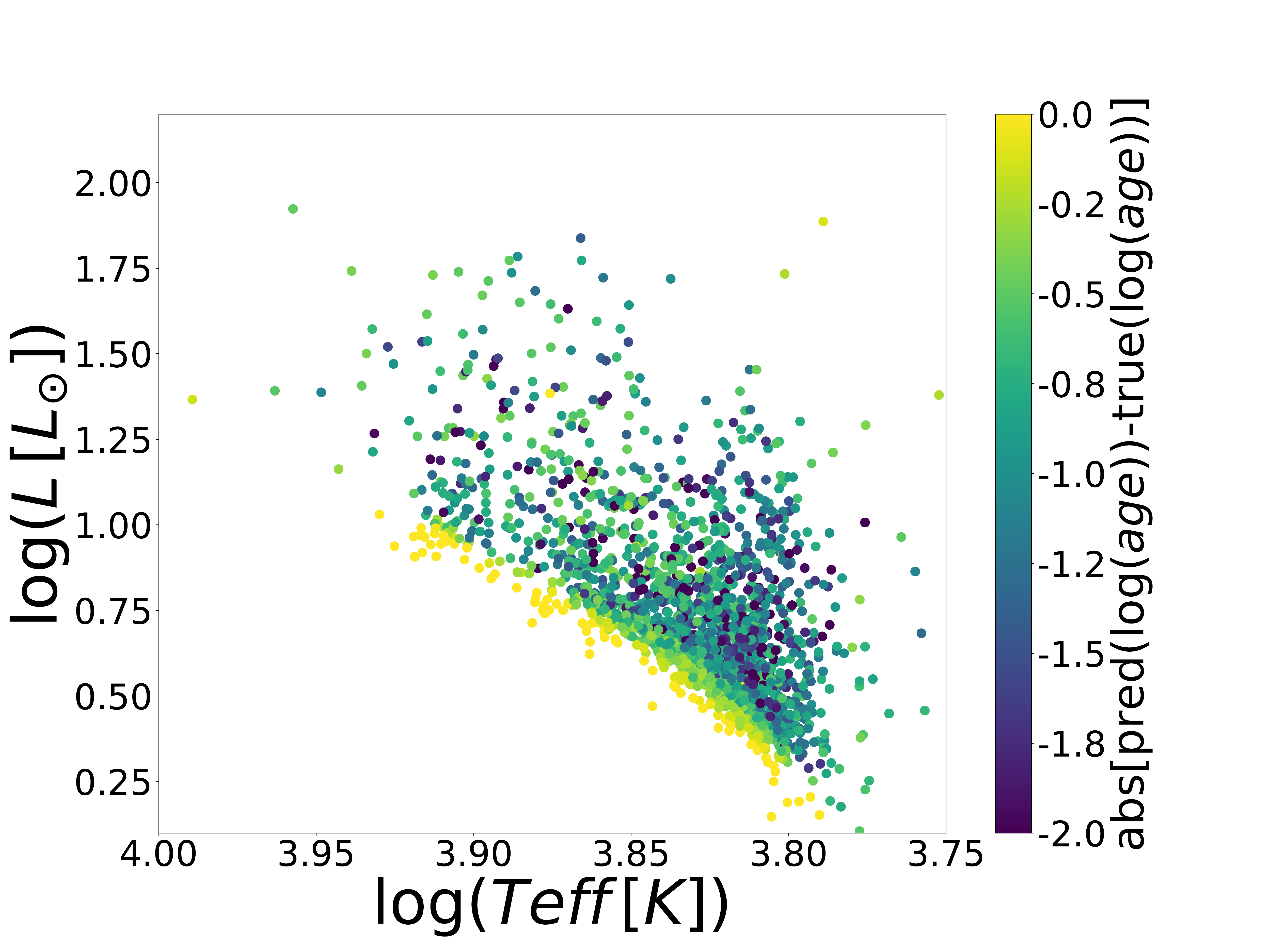}
\caption{$D\&H$ prediction errors on $log(age [yrs])$}
\label{fig:errors}
\end{figure}

Looking at these errors as a function of the input parameter space might be useful for understanding where the model performs well and where it needs improvement. In Figure~\ref{fig:errors} we show the log of the absolute errors, $\log(abs(pred(\log(age))-true(\log(age))))$ and $\log(abs(pred(\log(mass))-true(\log(mass))))$, as a function of the input parameter space.
We find that the model has issues predicting age to the left of the ZAMS where the turn-around point of the evolutionary tracks is located. This is not surprising since the D\&H dataset contains stars with observed parameters beyond that turning point. This should of course be prohibited from a stellar evolution perspective, but happens because of both observational error, and because the tracks are not ``perfect'' and real data can deviated from them. This situation shows a general bias coming from the lack of calibration of our model to real data. We will show in the next section a way to improve the model's performance.

In addition, we calculate the tension of each prediction with the reported values of age and mass. We find that 88.8\% of the D\&H age predictions (1660 out of 1869) have $\log{\tau} < 1$ and are, therefore, consistent with the reported ones (see middle panel of  Figure~\ref{fig:tension}), and 89.8\% (1679 out of 1869) for mass.

\subsection{Transfer Learning for Real Data}
\label{sec:TL}

So far, we have built and trained our model on synthetic data with no observational error. In reality, the observed stellar properties will differ from those predicted by the MESA tracks due to stellar evolution being an approximation, and, in addition, the observations of real stars will include an aleatory uncertainty that will place them at a commensurately offset point in the parameter space compared with their true locations. These two effects will result in systematic biases and less accurate predictions from our model, as reflected in Section~\ref{sec:val_gaia}. And this effect is emphasized by the fact that there are areas of the input parameter space where some predicted parameters change very quickly, for example, stellar age at the final stages of evolution. 

Machine learning methods work well under the assumption that the training and test data belong to the same distribution. In our case, the distributions are similar but not the same, limiting the model's performance. To mitigate this issue, we used Transfer Learning (see \citealt{pan2009survey} for a review). We used the network trained on synthetic data as a starting point and then fine-tune the weights using real data. This approach requires fewer real examples while enabling the use of a vast number of simulations. 

\subsection{Evaluation on Observed Data after Transfer Learning}
\label{sec:val_gaia_tl}

\begin{figure*}
\center
\includegraphics[trim = .1in 0.1in  0.1in 0in,clip, width =.7\linewidth]{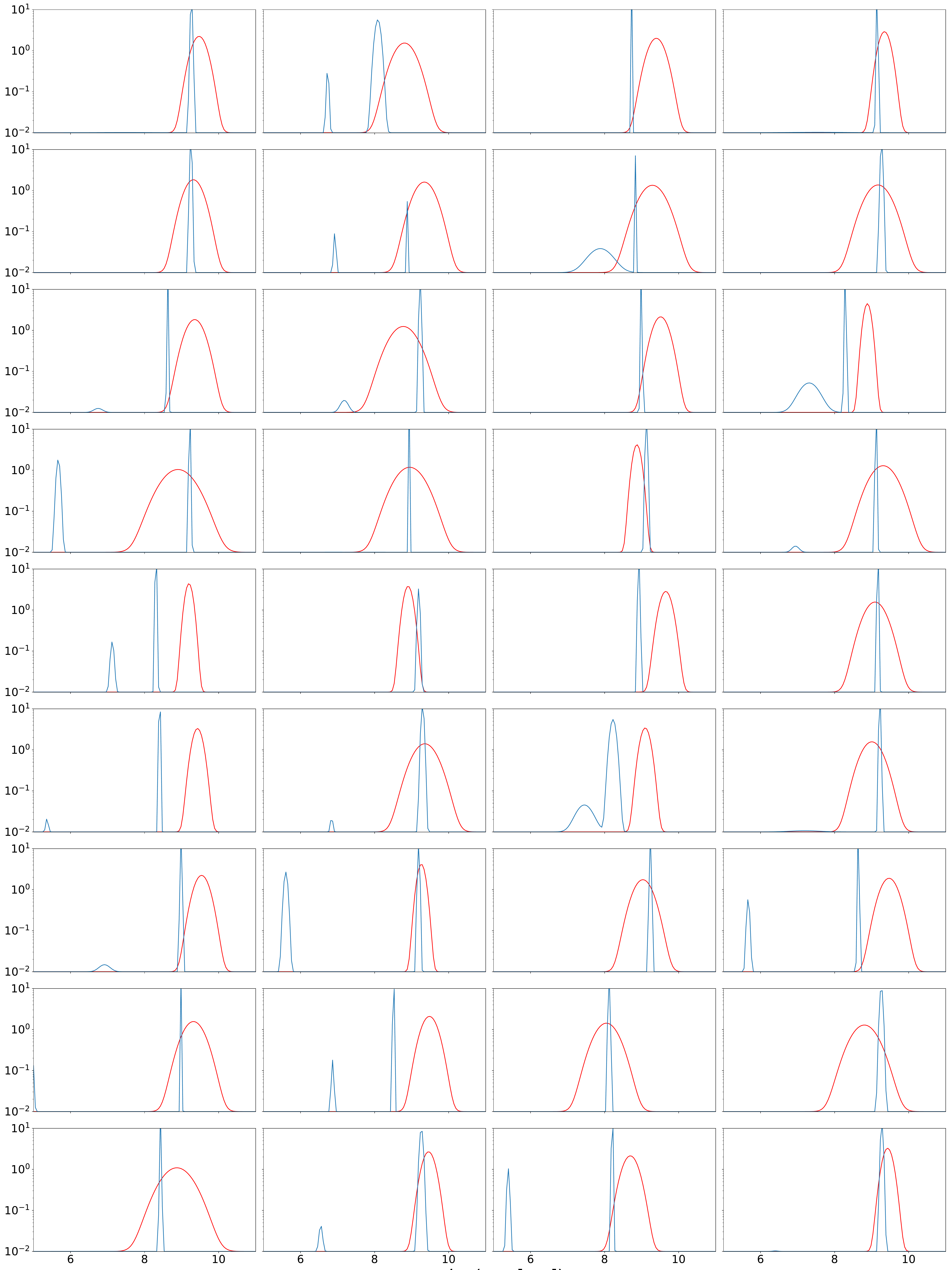}
\caption{Random selection of {\sc StelNet} posterior probability distributions for D\&H data on age (blue) and Gaussian representing observation with reported error (red) after TL.}
\label{fig:post_T&H}
\end{figure*}

We use transfer learning to re-train our network for 100 epochs on a hybrid of synthetic data (0.02\% of {\it MIST training data}) and observed data (20\% of the D\&H data) to improve the accuracy on real data.  We train in the same way as we did before, without freezing any part of the network. However, the relative probabilities between pre- and post-MS models should be reproduced better since they reflect the abundances of each stellar mass and age in the galaxy. In  Figure~\ref{fig:post_T&H} we show the posterior probability distribution of $\log(age)$ from our model for a randomly selected set of D\&H observations and the estimated age for each of them. In addition, we show the propagated observational error on age and mass estimates. We took 20\% of the observed dataset (D\&H) that has mass and age estimations, and we trained the network for a few (100) more epochs. That allows to retain the general trends of the model trained on synthetic data while calibrating on real data. 

\begin{figure}[htp]
\includegraphics[trim = .1in 0.in  0.1in .0in,clip, width =0.47\textwidth]{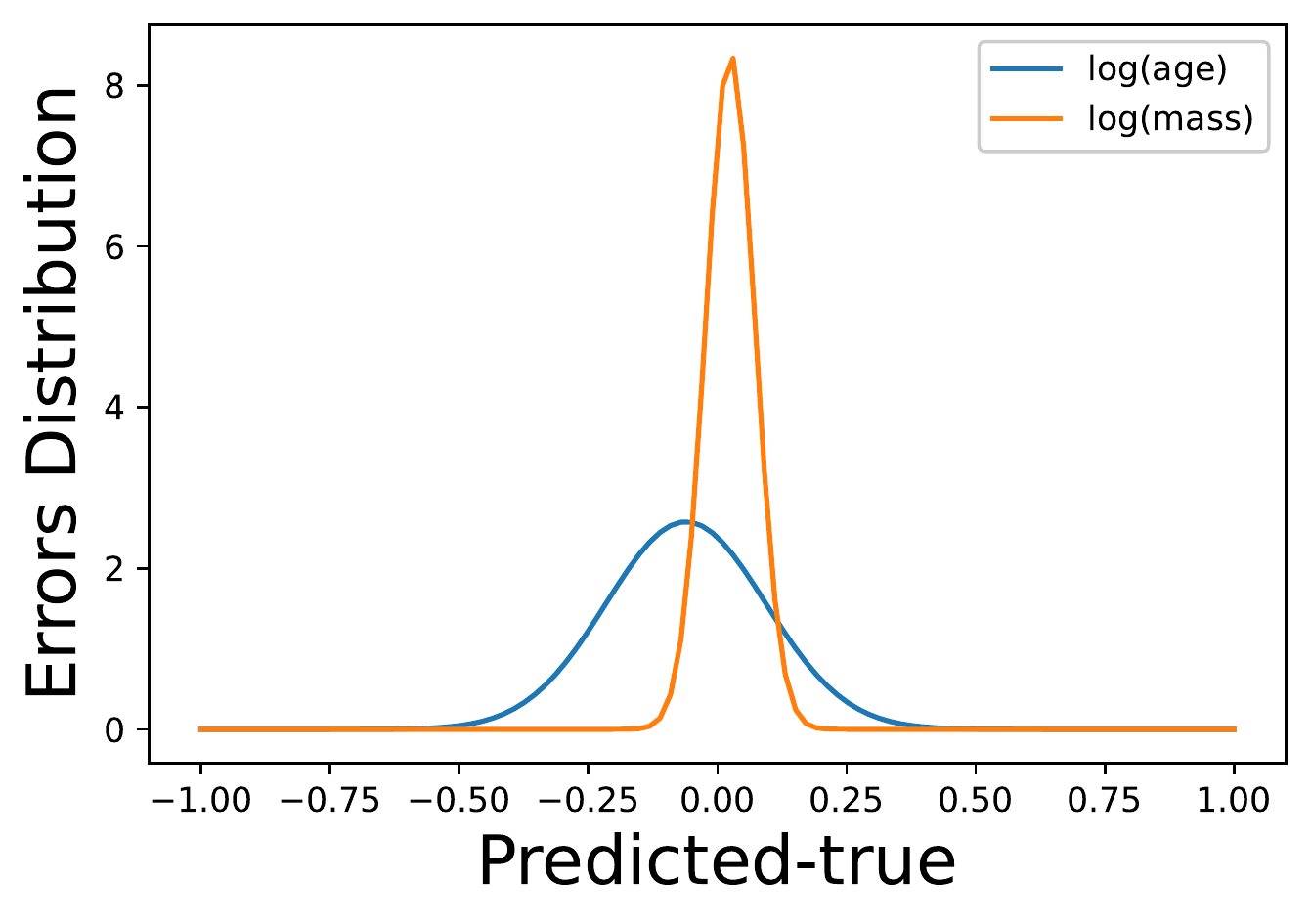}
\caption{The distributions of the $D\&H$ data prediction errors for age and mass after TL.}
\label{fig:stats_gaia_TL}
\end{figure}

\begin{figure}[htp]
\includegraphics[trim = 0.1in .1in  .5in .3in,clip, width =0.49\textwidth]{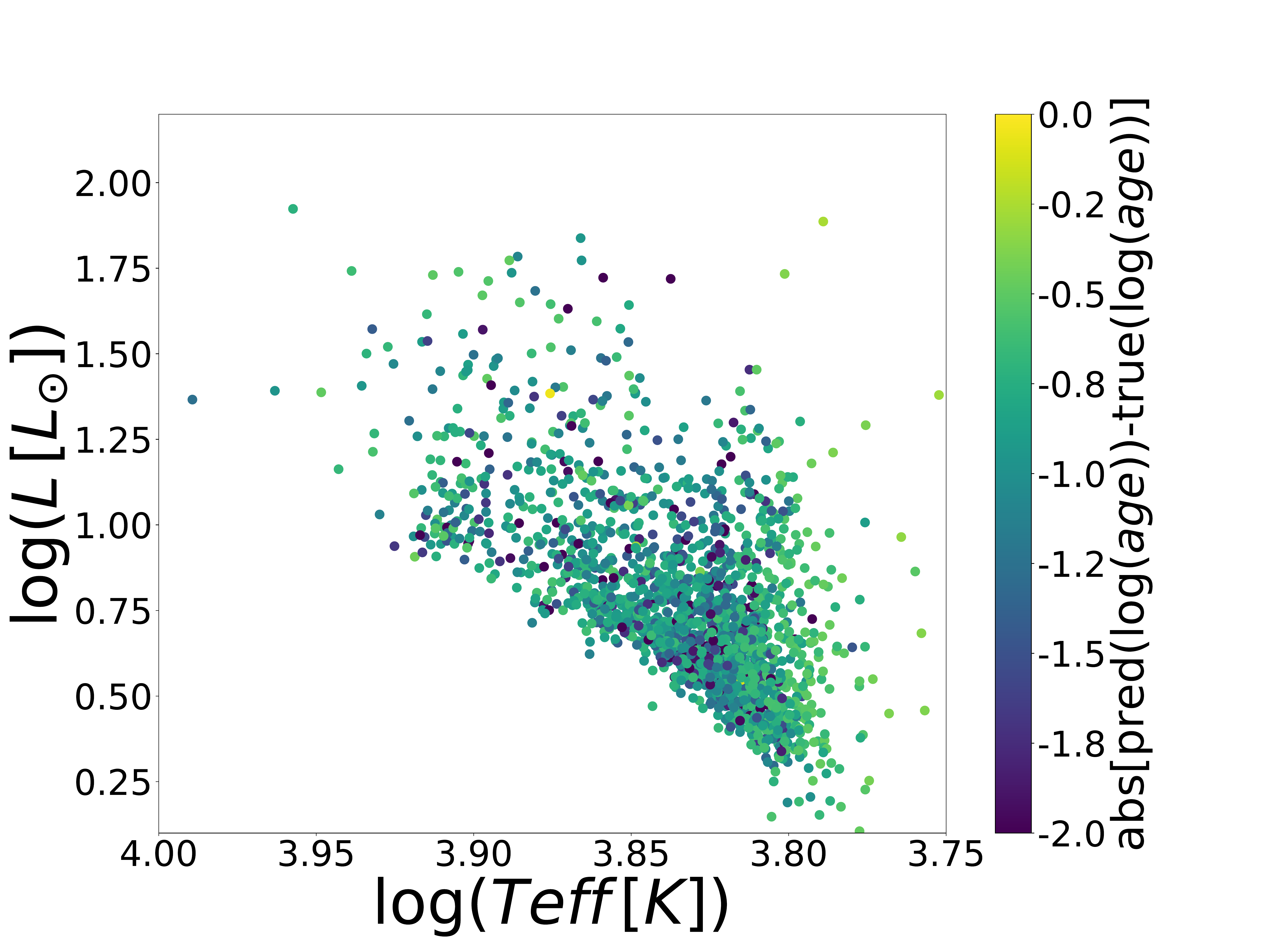}
\caption{$D\&H$ prediction errors on $log(age \, [yrs])$ after having performed transfer learning.}
\label{fig:errors_TL}
\end{figure}

We tested our new model on the D\&H dataset and found much better agreement than before performing transfer learning (see Figure~\ref{fig:post_T&H}).
To quantify the performance on this data set we, once again, calculate the distribution of errors for $\log(age)$ and $\log(mass)$, shown in Fig~\ref{fig:stats_gaia_TL}. The average error in $\log(age)$ predictions is now $\sim -0.06$ with a standard deviation of $\sim 0.15$, and the average error in $\log(mass)$ predictions is $\sim 0.02$ with a standard deviation of $\sim 0.047$. This represents a significant improvement in age predictions and their uncertainties, with no significant change in mass predictions. The reason behind this is that transfer learning has helped reduced systematic biases introduced by the choice of evolutionary tracks, which were affecting mainly the age predictions. This becomes clear when comparing Figure~\ref{fig:errors_TL} to Figure~\ref{fig:errors}.

In addition, after performing transfer learning training we find that 97.8\% (1827 out of 1869) of the D\&H age predictions have $\log{\tau}<1$ and, therefore, are consistent with the reported ones (see middle bottom Figure~\ref{fig:tension}), while 92\% of the mass estimates are similarly consistent. 

The reason for this improvement becomes more clear when looking, again, at the errors as a function of input parameter space after transfer learning (See Figure~\ref{fig:errors_TL}). We can see that the ill behavior to the left of the turn-around points (ZAMS) gets fixed or, at least, becomes much less noticeable. This has served as a calibration of our model to observed data.

\subsection{Evaluation on a triple system: $\alpha$ Centauri}

Multiple star systems typically are formed from a single cloud of gas and dust and, therefore, the stars in it all have the same age. 
An interesting test is to predict the age of a triple system of stars with different input parameters $T_{eff}$ and $L$, and compare the overlap in $\log(age)$ probabilities with the correct age. We do this test on the well-studied triple system $\alpha$~Centauri, 1.3 parsecs away, composed by $\alpha$~Centauri A, with $L= 1.519 \pm 0.018 \,L_{\odot}$ and $T_{eff} = 5790 \pm 30 \, K$ , $\alpha$~Centauri B, with $L= 0.5002 \pm 0.016 \,L_{\odot}$ and $T_{eff} = 5260 \pm 50 \, K$ \citep{Thevenin.etal:02}, and $\alpha$~Centauri C (also known as Proxima Centauri), with $L= 0.0017 \pm 0.0002 \,L_{\odot}$ and $T_{eff} = 3042 \pm 50 \, K$ \citep{Doyle.Butler:90, Segransan.etal:03}.

The $\alpha$~Cen system is generally considered coeval, with an age that has been well-constrained over the last two decades by
asteroseismic studies to be slightly older than the Sun, with various estimates ranging between 4.8 and 6.5~Gyr \citep{Thevenin.etal:02,Bazot.etal:12,Miglio.Montalban:05,Thoul.etal:03,Eggenberger.etal:04,Joyce.Chaboyer:18}. \citet{Mamajek.Hillenbrand:08} note that ages estimated from asteroseismology, ``gyrochronology" (rotation period), and chromospheric activity level are generally consistent.  
We adopt the error-weighted mean of the literature asteroseismology age results and their standard deviation of $5.73 \pm 0.73$~Gyr as a fiducial age for $\alpha$~Cen with which to compare our estimates.

The masses of the $\alpha$~Cen AB components have been tightly constrained from the orbital elements of the system as $M_A= 1.1055\pm 0.004$ and $M_B= 0.9373\pm 0.003$  \citep[e.g.][]{Kim:99,Kervella.etal:16,Kervella.etal:17}. The mass of Proxima~C has been estimated from mass-luminosity relations \citep[e.g.][]{Delfosse.etal:00} and 2MASS $K_s$ magnitude \citep{Mann.etal:15}. We adopt the 
latter value here, $M_C=0.1221 \pm 0.0022 M_\odot$.

\begin{figure}[htp]
\includegraphics[trim = .0in 0.in  0.1in .1in,clip, width =0.45\textwidth]{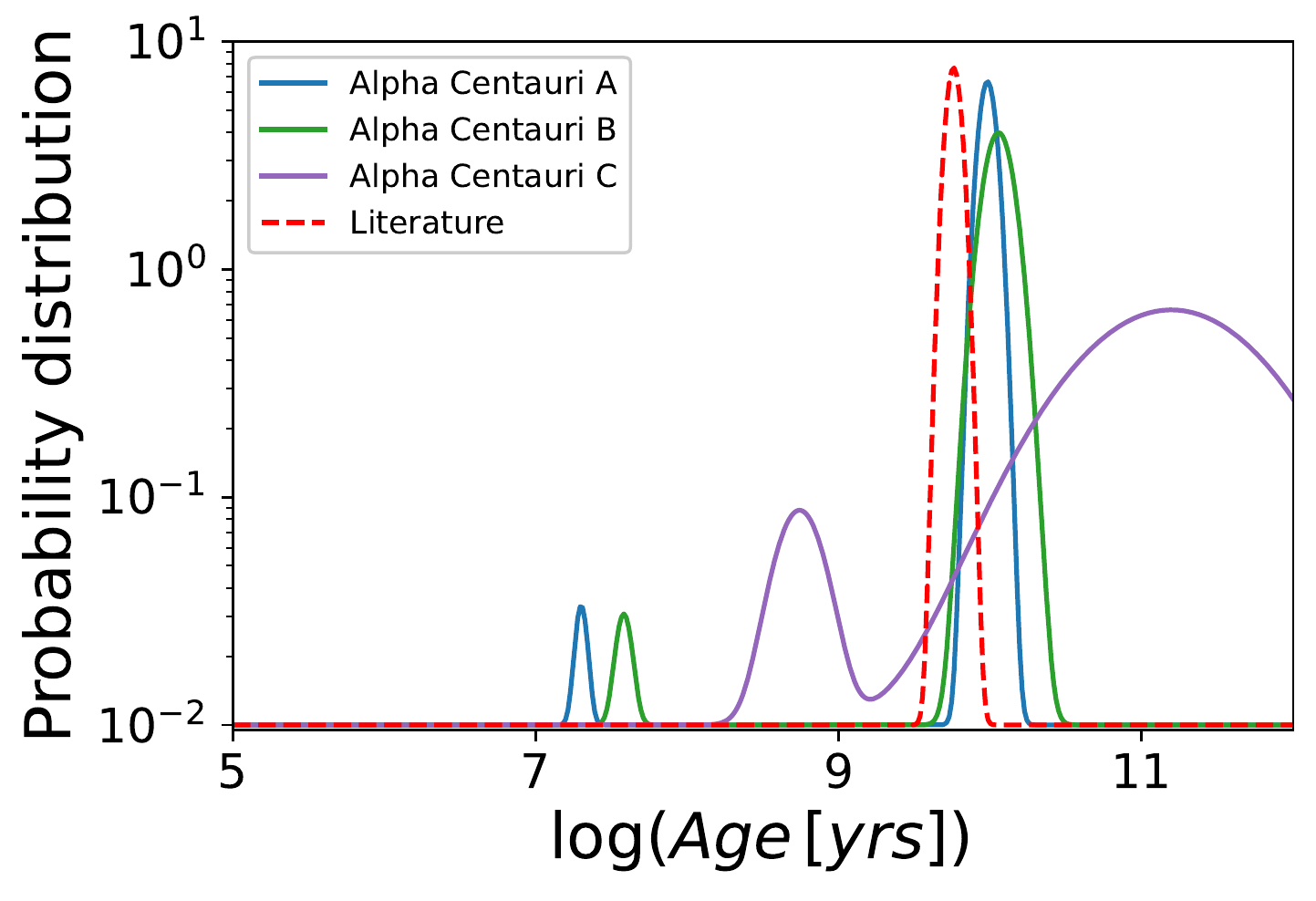}
\includegraphics[trim = .0in 0.in  0.1in .1in,clip, width =0.45
\textwidth]{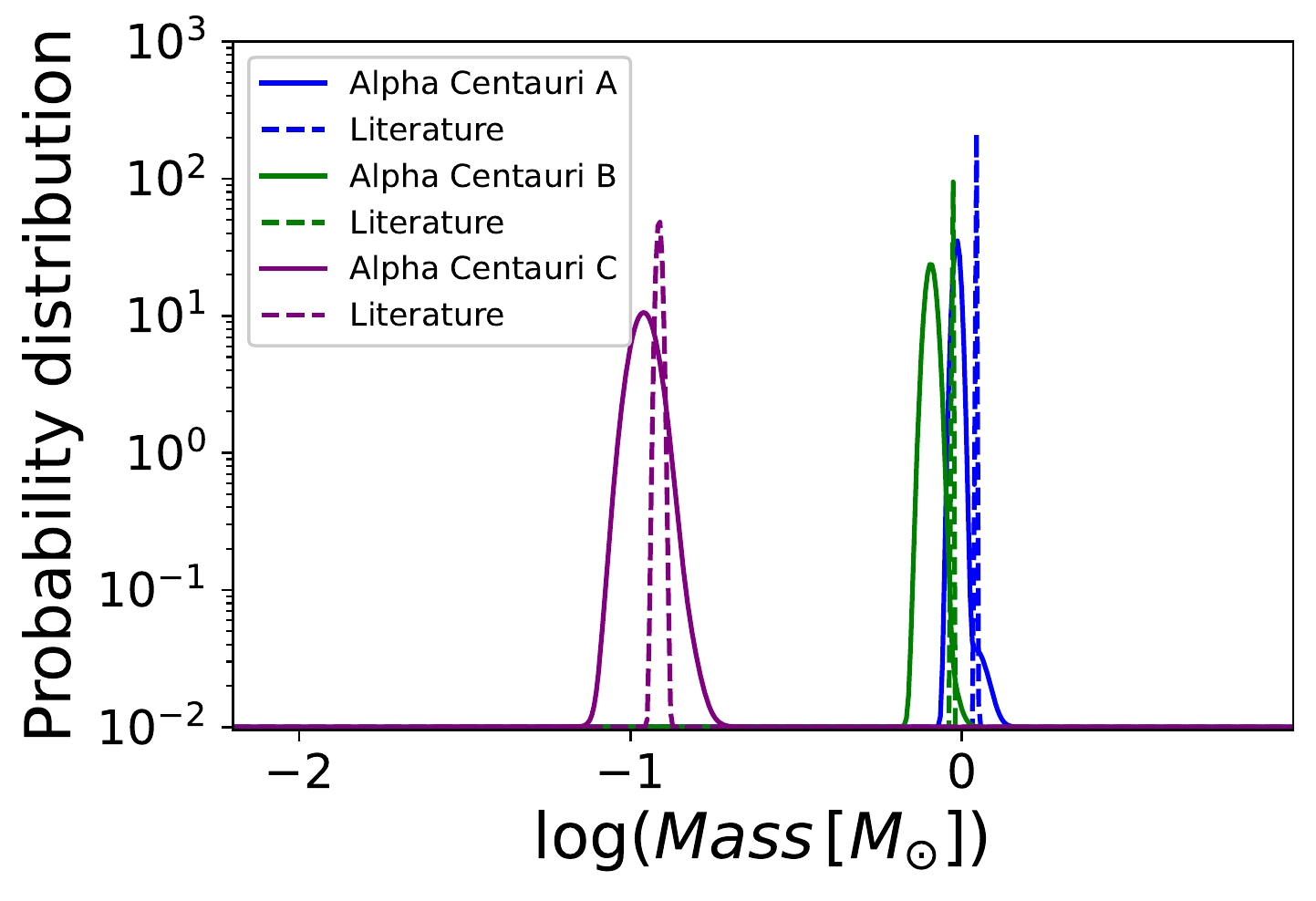}
\caption{Predicted posterior probability distributions for $\alpha$~Centauri stars A,B and C (solid lines) on age (top) and mass (bottom) and estimates from the literature (dashed lines) respectively.}
\label{fig:centauri}
\end{figure}

Comparison of {\sc StelNet} age and mass predictions for the $\alpha$~Cen components is illustrated in Figure~\ref{fig:centauri}. 
Our predictions have a significant overlap with the values from the literature, although are not in perfect agreement. The predicted probability of age is mildly consistent with the well-established age of the system (see top panel of Figure~\ref{fig:centauri}). For all three stars, the masses reported in the literature fall closer than 1.3 standard deviations away from {\sc StelNet}'s predictions (see top panel of Figure~\ref{fig:centauri}). 

Less than perfect agreement is expected in the case of $\alpha$~Cen, keeping in mind that {\sc StelNet} is at present limited to solar metallicity stars and $\alpha$ Cen's metallicity is higher than solar by 0.25 dex (or Z/X$= 0.04$
\citep[e.g.][]{Thevenin.etal:02,Porto_de_Mello.etal:08}.

Well constrained systems like $\alpha$~Cen will be used to calibrate our models in a follow-up paper using methods like transfer learning (see Section~\ref{sec:TL}).  

\subsection{The Sun}

Finally, we verify that {\sc StelNet} retrieves acceptable parameters for the Sun. The Sun presents an interesting case because its parameters are known to greater precision than for any other star. At the same time, reproducing {\em exactly} the observed solar characterstics is challenging for stellar evolution models because of the various approximations made in the models themselves, and because of imprecise knowledge of requisite input parameters such as element abundances and opacities. 

Using a canonical effective temperature of the Sun of 5771~K \citep{Heiter.etal:15}, and $1 L_\odot$, {\sc StelNet} retrieves an age of $4.9 \pm 0.7$~Gyr and a mass of $0.97 \pm 0.03 \, M_{\odot} $.

\section{Future work}

This work is a first step towards building a probabilistic model to accurately and quickly retrieve physical stellar parameters from observations. Here we have presented the ``vanilla'' model that predicts stellar mass and age for solar metallicity starts from $T_{eff}$ and $L$. 

The first planned enhancement will consist of including the evolutionary timescales for each step in the evolutionary models to fold into the probability of finding a star in a given evolutionary phase. Currently we use this idea but only to separate between the probability of being a pre-MS star versus one that has reached ZAMS.

We are currently training {\sc StelNet} to achieve a one-step model that retrieves the same parameters directly from observations. We will then train them directly on photometric data. The  
Gaia-ESO benchmark stars with effective temperature and luminosity determined through largely model-independent means \citep[e.g.][]{Jofre:16} would make a natural additional training set.

The next step is to include metallicity as a new output dimension. This will increase the degeneracies, which we plan to circumvent using the same hierarchical model strategy we presented here. 

Finally, we will include more predictors to improve the model. The natural one to constrain age predictions is rotation periods \citep[see, e.g.,][]{Angus.etal:19}. We will use {\it MIST} models that self-consistently include rotation in their stellar evolution \citep{Gossage.etal:20}. Other inputs that correlate with age are X-ray emission and space velocities. 

This model replaces the need to interpolate within tables of evolutionary tracks, dramatically reducing the data storage requirements. In addition, it is computationally efficient and allows for inference in very large datasets by circumventing the need of full probabilistic models like MCMC. In short, it can be useful as a quick estimate of parameters for large photometric datasets with epistemic errors and requires little data storage and computational run time. 

\section{Conclusions}
\label{sec:Results}

We have built two deterministic DNNs that convert $T_{eff}$ and $L$ to mass and age for solar metallicity stars, one for before and one for after they reach ZAMS.  Using bootstrapping, we have lifted these models into inference ones that quantify their epistemic uncertainties. We then built a hierarchical model to combine these models in a probabilistic way that considers the likelihood of observing a star of certain mass and age, and deals with the degeneracies of pre- and post-ZAMS stars, inherent to the color-magnitude diagram parameter space. Our final model, {\sc StelNet}, takes as inputs $T_{eff}$ and $L$ and provides realistic posterior probability distributions on the physical parameters mass and age, typically bi-peaked due to the degeneracy between pre- and post-ZAMS stars. 

We have trained this model on a fraction of a {\it MIST} isochrones 
dataset and have validated it on the remaining fraction. We find that 98.3\% of predictions are consistent with our predictions.

In addition, we validated our model against an observed D\&H dataset with reliable mass and age estimates. We find that that 88.8\% of predictions are consistent with our predictions, which is a good agreement considering that our only input features are $T_{eff}$ and $L$, these observations have errors, and that the synthetic data our model was built on does not exactly reflect the evolution of real stars.  

In order to boost the performance of our model on real data, we used transfer learning. We used the model trained on synthetic data as a starting point and used a small set of the observed dataset D\&H together with a small set of {\it MIST} data to calibrate our model. After this step we find that our performance increased, significantly, to 97.8\% of predictions being consistent with the observed data.

Finally, we tested our model on the well-studied triple system $\alpha$ Cen. The three stars in this system have different input parameters $t_{eff}$s and $L$s, but their ages are the same. We find that the age of the system reported in the literature, 5.73 Gyrs, is not far from the range of largest overlap in age posterior probability prediction for age (6-10 Gyrs), which is good agreement if we take into account that $\alpha$ Cen is metal rich and {\sc StelNet} was designed for solar metallicity stars.

\acknowledgments

CG thanks the research group at IACS, Javier Machin and Nicholas J. Wright for very helpful discussions. JJD  was  funded  by  NASA  contract NAS8-03060 to the {\it Chandra} X-ray Center and thanks the Director, Pat Slane, and the CXC science team for continuing advice and support.  We thank anonymous referee for a very constructive report that has helped improve this work.

\bibliographystyle{apj.bst}
\bibliography{StelNet.bib}

\end{document}